\documentclass{aa}  

\usepackage{graphicx}
\usepackage{hyperref} 
\usepackage{txfonts}
\usepackage{float}
\usepackage{lipsum}
\usepackage{subcaption}         
\usepackage{lscape}             
\usepackage{placeins}           
                                
\begin{document}


 \title{Distinct Population of Wolf-Rayet Stars in the Low-Metallicity Galaxy NGC~6822\thanks{Based on observations made with the NASA/ESA {\em Hubble} Space Telescope, 
   obtained from the data archive at the Space Telescope Science Institute. STScI is operated by the Association of Universities for Research in Astronomy, 
   Inc.\ under NASA contract NAS 5-26555. These observations are associated with the GO program 17732. Based on observations collected at the European Organisation for Astronomical Research in the Southern Hemisphere under ESO programme 109.2326.001.}}

   \subtitle{Quantitative analysis of optical and UV spectra of the complete  sample}


   \author{R. Trigg\inst{1}\thanks{rtrigg@astro.physik.uni-potsdam.de}
          \and
          L.\,M. Oskinova\inst{1}
          \and
          W.-R. Hamann\inst{1}
          \and
          H. Todt\inst{1}
          \and
          D. Pauli\inst{2}
          \and
          T. Shenar\inst{3}
          \and
          A.A.C. Sander\inst{4}
          \and
          M. Chatzis\inst{1}
          \and
          A. Mang\inst{1}
          \and
          R.R. Lefever\inst{4}
          \and
          V. Ramachandran\inst{4}
          \and
          D. Massa\inst{5}}

   \institute{Institut für Physik und Astronomie, Universität Potsdam, Karl-Liebknecht-Str. 24/25, 14476 Potsdam, Germany
    \and{Institute of Astronomy, KU Leuven, Celestijnenlaan 200D, 3001 Leuven, Belgium}
    \and{The School of Physics and Astronomy, Tel Aviv University, Tel Aviv 6997801, Israel}
    \and{Zentrum für Astronomie der Universität Heidelberg, Astronomisches Rechen-Institut, Mönchhofstr. 12-14, 69120 Heidelberg, Germany}
    \and{Space Science Institute (SSI), 4750 Walnut Street, Suite 205, Boulder, CO 80301, USA}}

   \date{Received ??}

  \abstract
{Wolf-Rayet (WR) stars are hydrogen-depleted, highly evolved massive stars characterised by emission line spectra originating in their powerful stellar winds. WR stars are particularly rare in low-metallicity galaxies and remain poorly understood. Detailed studies of the presumably complete population of WR stars in the nearby Small Magellanic Cloud (SMC) galaxy, with metallicity $\sim 20$\%\, led to perplexing results and called for new investigations of WR stars in low metallicity galaxies.}
{The nearby galaxy NGC~6822 has a similar metallicity to the SMC, and hence provides an excellent laboratory for a comparative study. We aim to derive the fundamental stellar and wind parameters of each known WR star in NGC~6822 by means of quantitative spectroscopy. The results are compared with the WR population of the SMC to further our understanding of the final evolutionary stages of massive stars at low metallicity.}
{New optical and UV spectroscopic observations have been obtained with the Very Large Telescope (VLT) and Hubble Space Telescope (HST). We employ the Potsdam Wolf-Rayet (PoWR) model atmosphere code for our analysis, consistently fitting synthetic spectra to the observed spectral energy distribution, and optical and UV spectra.}
{We spectroscopically confirm that all four stars in our sample are of WR type and belong to the nitrogen sequence with early subtypes (WN3--WN6). The spectra of all stars show contributions from a secondary star and are therefore analysed as composite. Only one star in our sample has a high fraction of hydrogen in its atmosphere, as typical for the WN population in the SMC. Two other stars in our sample are hydrogen-free. Finally, the spectrum and parameters of the fourth object closely resemble those of the qWR-type strongly magnetic merger-product star in the HD\,45166 binary.} 
{The stellar properties of our sample are quite diverse. All stars show indications of binarity. We have identified the first H-free WN-type stars at metallicities $\sim 1/7 Z_\odot$. According to evolutionary calculations, the WR stars in NGC~6822 stem from progenitors with initial masses below $40\,M_\odot$. We report a discovery of a low luminosity stripped qWR star in NGC~6822. We conclude that the WR star population in NGC~6822 and the SMC are distinct despite similar metallicities of their host galaxies.  
}


   \maketitle
\nolinenumbers
\section{Introduction}\label{Introduction}

Wolf-Rayet (WR) stars are highly evolved massive stars that have lost a large portion of their outer H-rich layers \citep[see][for a review]{WRBASICS_Crowther2007}. The strong winds of WR stars cause their characteristic spectra, which are dominated by broad emission lines of metals and helium. WR stars are observed in several sub-types that are indicative of their surface composition, temperature, and evolutionary stage. The broadest classification scheme distinguishes between the nitrogen (WN), carbon (WC), and oxygen (WO) sub-types.

About 40\% of WR stars are confirmed binaries \citep[][]{BINARIES_vanderHucht2001, Dsilva2023, Deshmukh2024}, and this is seemingly consistent across metallicities \citep[][]{WRLMC_hainich2014, M31M33_NeugentMassey2014, WRSMC_Shenar2016}. The mass-loss rates ($\dot{M}$) of stars with radiation-driven winds depend on metallicity ($Z$). Therefore, at low metallicity, only the most luminous stars can shed their hydrogen envelopes by stellar winds, and binary stripping should be the prevailing mechanism in the formation of WR stars \citep[e.g.][]{BINARIES_Eldridge2022, Pauli2022}.

The SMC, with a metal content around one fifth of the solar value \citep{SMCMET_Korn2000, SMCMET_Hunter2007}, is so far the lowest-$Z$ galaxy where a complete sample of WR stars  has been systematically investigated \citep[12 objects,][]{SMCSURVEY_Massey2014, SMCSURVEY_Neugent2018}. Intensive monitoring revealed that 5 out of these 12 WR stars (40\%) are confirmed binaries \citep{SMCBIN_Foellmi2003,MASSLOSS_Schootemeijer2024}. Except of one WO-type star, all other  one WR stars in the SMC are very luminous and have a WN sub-type.  What is particularly striking is that all WN stars in the SMC contain significant fractions of hydrogen in their atmospheres \citep{SMCWR_Martins2009, WRSMC_Hainich2015, WRSMC_Shenar2016}. In this respect, they are fundamentally different from the WR populations in the Galaxy and the Large Magellanic Cloud (LMC). At the higher metallicities of these galaxies, most WN stars hotter than the zero-age main sequence (ZAMS) are hydrogen-free \citep{WRSURVEY_Hamann2006,WRLMC_hainich2014}.

The origin and evolution of the WN stars in the SMC remain enigmatic, calling for new evolutionary scenarios \citep{Li+2024, TRACKS_pauli2026}. The positions of the WR stars in the Hertzsprung-Russel diagram (HRD) together with their high hydrogen content can in principle be explained with stellar evolution models that assume quasi-chemically homogeneous evolution, due to strong internal mixing \citep[][]{brott+2011, WRQCHE_Boco2025,SMCWR_Martins2009, WRSMC_Hainich2015}. However, this scenario has been debated \citep{SMCWR_Schootemeijer2018, SMCWR_Ramachandran2019, WRSMC_Shenar2016, SMCWR_Aguilera-Dena2022, EVO_Pauli2023, EVO_Martins2023}. It is therefore of paramount importance to investigate whether the conspicuous properties of WR stars in the SMC reflect specific evolutionary pathways of massive stars in low metallicity galaxies. This task can be achieved only by comparative study of the WR population in another galaxy with a comparable metallicity.

\begin{figure}
	\begin{center}	
        \includegraphics[width=0.9\linewidth]{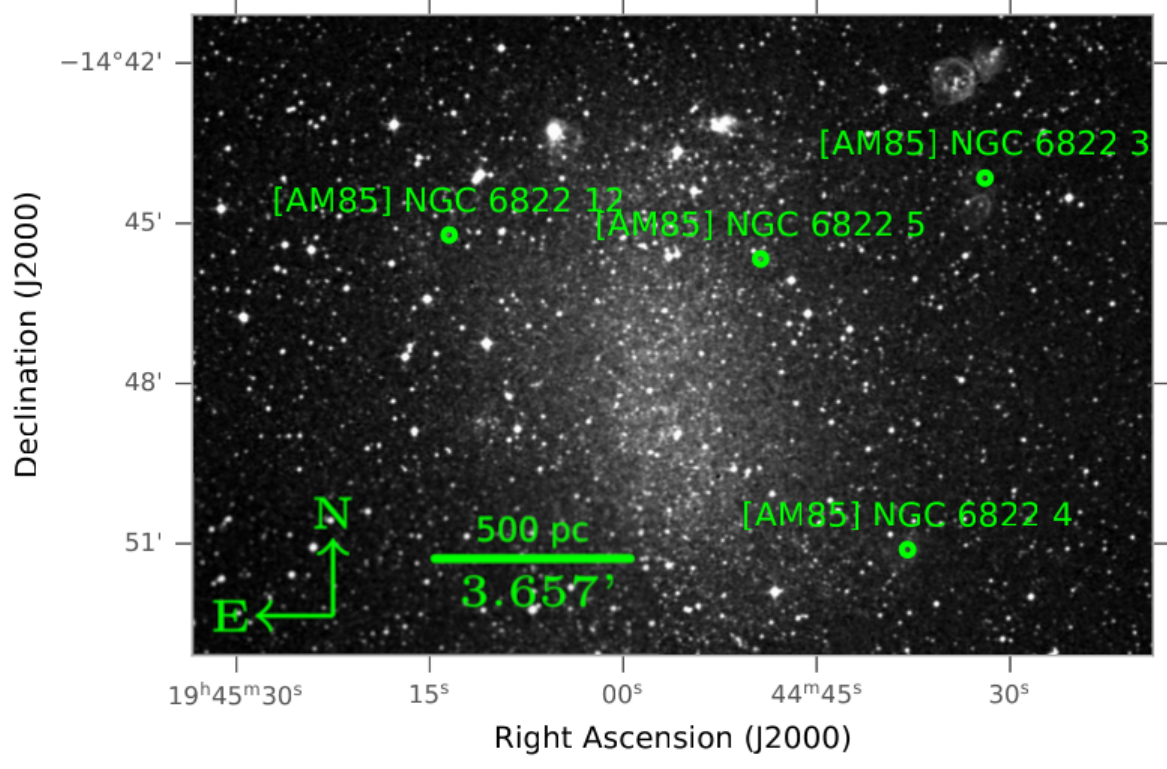}
	\end{center}
    \caption{A combined image of NGC~6822 (DSS red and DSS blue, in grey scale). The image was made with \href{https://alasky.u-strasbg.fr/hips-image-services/hips2fits}{hips2fits} \citep{Hips2Fits}. The green circles mark our sample of WR stars. The labels give object names according to Table\,1.}
    \label{NGC6822distances}
\end{figure}

In this paper, we present the first spectroscopic  WR population in  NGC~6822, also known as Barnard's Galaxy. It is a nearby ($d = 490 \pm 40$~kpc), isolated dwarf irregular galaxy with a similar stellar mass and recent star formation rate to the SMC \citep[e.g.][]{SFH_Fusco2014, NGC_Khatamsaz2024}.  Most importantly, the metallicities of NGC~6822 and the SMC are also similar -- the oxygen abundance is higher in NGC~6822 compared to the SMC, whilst the iron abundance is similar or even lower than in the SMC \citep{SMCMET_Pilyugin2001, NGCMET_Hernandez-Martinez2009,Patrick2015}. The WR sample in NGC~6822 is considered to be complete, containing four objects \citep{NGCWR_Armandroff1985A, NGCWR_Armandroff1991}. Figure~\ref{NGC6822distances} shows the location of the WR stars within the galaxy. 

The paper is organised as follows. In Section\,2, we describe the observations, and in Sect.\,3 the stellar atmosphere model used in the analyses. The results and derived stellar and wind parameters are detailed in Sect.\,4. A discussion on stellar properties and evolution is presented in Sect.\,5, and the conclusions are summarised in Sect.\,6. 

\begin{table}\centering
	\caption{Sample of WR stars in the galaxy NGC~6822}
		\begin{tabular}{ccccc}
			\hline\hline
			Name [AM85] 	 & RA  & DEC  & Spectral \\
            NGC~6822 \#  & J2000 & J2000 & Type \\
			\hline
            3 & 19:44:31.97 & -14:44:09.33 & WN6 \\
            4 & 19:44:37.96 & -14:51:06.81 & WN3 \\
            5 & 19:44:49.35 & -14:45:40.27 & WN3 \\
            12 & 19:45:13.49 & -14:45:13.28 & WN3-4 \\
			\hline
            
		\end{tabular}	
        \tablefoot{First column: the star identifiers increase with increasing right ascension. Second and third columns: coordinates from \cite{NGCWR_Armandroff1985A}. Fourth column: spectral types from this work, following criteria in Table\,2 in \citet{WRCLASS_vanderhucht2001}.}
        \vspace*{-3mm}
	\label{tab:Star_names}
\end{table}

\section{Observations}\label{Observations}

\subsection{Optical and UV spectroscopy}\label{Obs_Optical_section}

Optical spectra were taken using the FORS~2 spectrograph \citep{FORS2_Appenzeller1998} mounted on the UT~1 ESO VLT (Programme ID:109.2326.001, PI: L. Oskinova) -- see Table~\ref{Obs_optical}. The spectra cover a range between 4625\,\AA\, and 7149\,\AA\, with a resolving power of about $R=2100$. The data reduction was performed with the ESOREFLEX \citep[][]{ESOREFLEX_Freduling2013} pipeline \href{https://www.eso.org/sci/software/pipelines/index.html\#reflex_workflows}{source kit 5.6.2} using default parameters. For each star, the spectra from the two grisms were merged and then normalised, setting the continuum points by hand. An exposure time of 2800~s was chosen to yield a signal-to-noise ratio of 30, for a seeing of 0\farcs8 full width half maximum.

UV spectra were obtained using the Cosmic Origin Spectrograph \citep[COS,][]{COS_Green2012} onboard the HST (GO program 17732, PI: L. Oskinova). The low-resolution ($R\sim2000$) G140L grating centred at 800\,\AA\ was used. The log of observations is shown in Table~\ref{tab:Obs_UV}.

Special care was taken when observing [AM85]~NGC~6822~5 because there is a second star at a separation of 1\farcs7, i.e.\ within the COS aperture. In order to minimise possible contamination, the orientation angle was restricted.
 
\subsection{Optical photometry}\label{Obs_Photometry}

We have extracted photometric magnitudes from HST WFPC2 images available in the MAST archive with the filters F170W, F225W, F336W, F439W, F555W, and F814W, using the \texttt{DAOphot} task in \texttt{IRAF} (Table~\ref{Stellar_parameters_all}). We also include Gaia G, $G_{BP}$, and $G_{RP}$ photometry from Gaia Data Release 3 \citep{GAIADR3_2016, GAIADR3_2023}. For \#12, we use additional photometry from \cite{NGCWR_Bianchi2001} (F170W, F225W, F336W, F439W, F555W, and F814W filters).

\subsection{X-ray non-detection}

NGC~6822 was observed by the \textit{Chandra X-ray Observatory} (PI: Tennant; ObsID 2925) with the ACIS-I array for an exposure of $28$~ks. The data were reprocessed applying the latest calibrations with version 4.17 of the Chandra Interactive Analysis of Observations (CIAO)\footnote{\url{https://cxc.cfa.harvard.edu/ciao/}}. No X-ray counterparts are detected at the positions of the four WR stars. For a $1~\sigma$ detection with a $5$~\% false positive probability, we derive upper limits of $L^{0.5-7~\text{keV}}_{\text{lim}} \lesssim 8 \times 10^{34}~\text{erg~s}^{-1}$. These do not rule out any of the WR stars being colliding wind binaries, which have typical X-ray luminosities of $\lesssim 10^{33}~\text{erg~s}^{-1}$ \citep{Oskinova2005}.

\section{Stellar Atmosphere Modelling}\label{Stellar Atmosphere Modelling}

The spectral analysis is performed using the state-of-the-art stellar atmosphere code PoWR which is suitable for the analysis of hot stars with winds. Local thermodynamic equilibrium is not assumed (non-LTE). The PoWR code solves the radiative transfer equation for a spherically expanding atmosphere and the statistical equilibrium equations simultaneously under the constraint of energy conservation \citep{POWR_Grafner2002, POWR_Hamann2003, POWR_Hamann2004, POWR_Sander2015, POWR_Todt2015}.

The basic input parameters for a model are the luminosity ($L_\ast$) and stellar temperature ($T_\ast$), as well as the mass-loss rate ($\dot{M}$), terminal wind velocity ($\varv_\infty$), and clumping factor ($D$) for the wind. The inner boundary of the model atmospheres is located at a Rosseland continuum optical depth of $\tau_{\text{Ross}}$ = 20, which per definition corresponds to the stellar radius $R_\ast$. The stellar temperature $T_\ast$ is related to this radius by the Stefan-Boltzmann law $L_\ast = 4 \pi \sigma R_*^2 T_*^4~$. The stellar temperature thus represents an effective temperature at $\tau_{\text{Ross}}$ = 20. The abundances must also be specified, and here we include H, He, C, N, O, and the Fe group elements. The iron group elements are treated with the 'superlevel' approach as described in \citet{POWR_Grafner2002}

In the subsonic region of the stellar atmosphere, a velocity field is defined such that a hydrostatic density stratification is approached \citep{POWR_Sander2015}. The velocity field in the supersonic part of the stellar wind is commonly prescribed by a $\beta$-law \citep{BETA_Castor1979, BETA_Pauldrach1986}, or a double-$\beta$ law \citep{BETA_Hillier1999, BETA_Grafener2005}. In this work, in the case of star \#12 and the single-star models for all stars (see App.~\ref{App:models}), we use a tabulated wind velocity as typically obtained by hydrodynamically consistent WR models calculated by \citet{VELO_Lefever2023,lefever2026} with the corresponding version of the PoWR code. This velocity table is shifted and scaled as required, and is shown in Fig.~\ref{Velocity field} in Appendix~\ref{App:models}.

The clumping factor $D$ describes small-scale, optically thin density inhomogeneities. The density in the clumps is enhanced compared to a homogeneous wind of the same $\dot{M}$ by a factor $D$. The mass-loss rate, as obtained from recombination lines, scales with $D^{-1/2}$ \citep[e.g.][]{CLUMPING_Hamann1998}.

In the non-LTE iteration in the co-moving frame, the line opacities and emissivity profiles are treated as Gaussians with a width following from a constant Doppler velocity $\varv_{\text{Dop}} = 100$\,km\,s\textsuperscript{-1}. In the formal integral for the calculation of the emergent spectrum, the Doppler velocity is split into the depth-dependent thermal velocity and a “microturbulence velocity” $\xi(r)$. Pressure broadening and rotational broadening can also be taken into account in the formal integral.

The fitting procedure consists of two parts, namely the fit of the spectral energy distribution (SED), and detailed fitting of the normalised line spectra. For the normalisation of the flux-calibrated UV spectra, we divide it by the reddened theoretical continuum. 

We employ the PoWR SMC WR grid models \citep[][]{POWR_Hamann2004, POWR_Todt2015} as a starting point, choosing the closest grid model by eye. The luminosity of each WR star (and its possible companion) and the colour excess $E(B-V)$ are obtained by fitting the model SED to the photometry and flux-calibrated UV spectra. Where Gaia photometry differs from our photometry (see Sect.~\ref{Obs_Photometry}), we aim to achieve a compromise (ignoring clear outliers).

Interstellar absorption is applied to the model flux. The reddening encompasses contributions from NGC~6822 and the Galactic foreground. For the Galactic component, we use the reddening law published by \citet{REDDENING_Seaton1979} and a colour excess of $E(B-V)$= 0.21~mag \citep{REDDENING_Schlafly2011}. The colour excess for the internal component from NGC~6822 is varied as a free parameter. The extinction for NGC~6822 appears quite anomalous and is poorly reproduced by standard reddening laws. A steep drop-off is observed towards the blue end of the UV spectra, whilst the red end of the UV spectra appear flatter than in the models. As the best compromise, we adopt the reddening law from \citet{REDDENING_Cardelli1989} with $R_V = 1.0$, but the SED fit is still imperfect for some stars (in particular for stars \#3 and \#5). In these cases we slightly re-normalise the UV spectrum by hand. Ly$\alpha$ and the interstellar absorption lines Si\,{\scshape ii} $\lambda$1260\,\AA, Si\,{\scshape ii} $\lambda$1527\,\AA, O\,{\scshape i} $\lambda\lambda$ 1302, 1304\,\AA, C\,{\scshape ii} $\lambda\lambda$1335, 1336\,\AA, and C\,{\scshape iv} $\lambda\lambda$ 1548, 1550\,\AA\, are accounted for in the theoretical SEDs and in the normalised spectra according to column densities corresponding to the reddening value \citep{HCOLMW_Groenewegen1989}.

Due to the long exposure, the UV spectra are contaminated by strong geocoronal Ly$\alpha$ emission. In order to examine its impact on neighbouring stellar lines, we model this line with a Voigt profile adjusted to the observation by hand. Additional geocoronal emission lines are also visible in the UV spectra, but not modelled.

In spectra of all our sample WR stars, we find signatures of a binary companion (see details in Sect.\,\ref{sec:results}). To determine the relative contributions of the binary components, we use the absorption features which are clearly associated with the O-type star. We model the companion spectra using the SMC-OB-Vd3 PoWR grid models\footnote{\url{https://www.astro.physik.uni-potsdam.de/PoWR}} (Pauli et\,al. submitted). The errors on the OB-star model parameters are comparable with the size of the grid cell ($\Delta\log{g}=0.2, \Delta T_\ast=10$\,kK). It is assumed that the companion rotates with $v\sin{i}=100$\,km\,s$^{-1}$ unless indicated otherwise. Ideally, the composite spectrum could be disentangled into its constituent spectra using observations taken at different phases, but our data do not enable this.

We systematically adjust the model parameters and conduct visual inspections until no significant improvement to the fit of prominent lines can be achieved (at a few percent level). The stellar temperature is inferred from the line ratios of ions belonging to the same element, and the mass-loss rate is adjusted such that the observed strengths of the emission lines are reproduced by the model. The terminal velocity is derived from the width of the emission lines and the blue edge of the P~Cygni line profiles of UV resonance lines like C\,{\scshape iv}~$\lambda\lambda$~1548,1551\,\AA, when available. Density contrasts of $D=4$ and $D=10$ are used for the WR and OB stars respectively, and adjusted where necessary. 

Abundances are determined from the strengths of lines belonging to the respective elements. The hydrogen content is estimated by comparing the hydrogen/helium blends H$\beta$ $+$ He\,{\scshape ii} 8-4 and H$\alpha$ $+$  He\,{\scshape ii} 6-4 with the unblended line He\,{\scshape ii} 7-4 at $\lambda$5412\,\AA.

We examine the sensitivity of the fits to changes in the stellar parameters to estimate uncertainties. These include errors on $T_\ast$, $\log(L)$, $\dot{M}$, $\varv_{\infty}$ and the abundances -- error propagation is used for the remaining parameters. Stellar masses are estimated for WR stars from mass-luminosity relations by \citet{MASSLUM_Grafener2011}. We use the relation for chemically homogeneous core He-burning stars when $X_\text{H} \leq 0.05$, and for H-burning otherwise.

\section{Results}\label{Results}
\label{sec:results}

In this section, we present the spectral analyses of the sample stars. The spectral classification is according to the criteria in \citet{WRCLASS_vanderhucht2001}. The derived stellar and wind parameters are compiled in Table~\ref{Stellar_parameters_all}. 

\subsection{NGC 6822 12}\label{WR12_analysis}

\begin{figure*}
	\begin{center}	
		\includegraphics[width=0.7\linewidth]{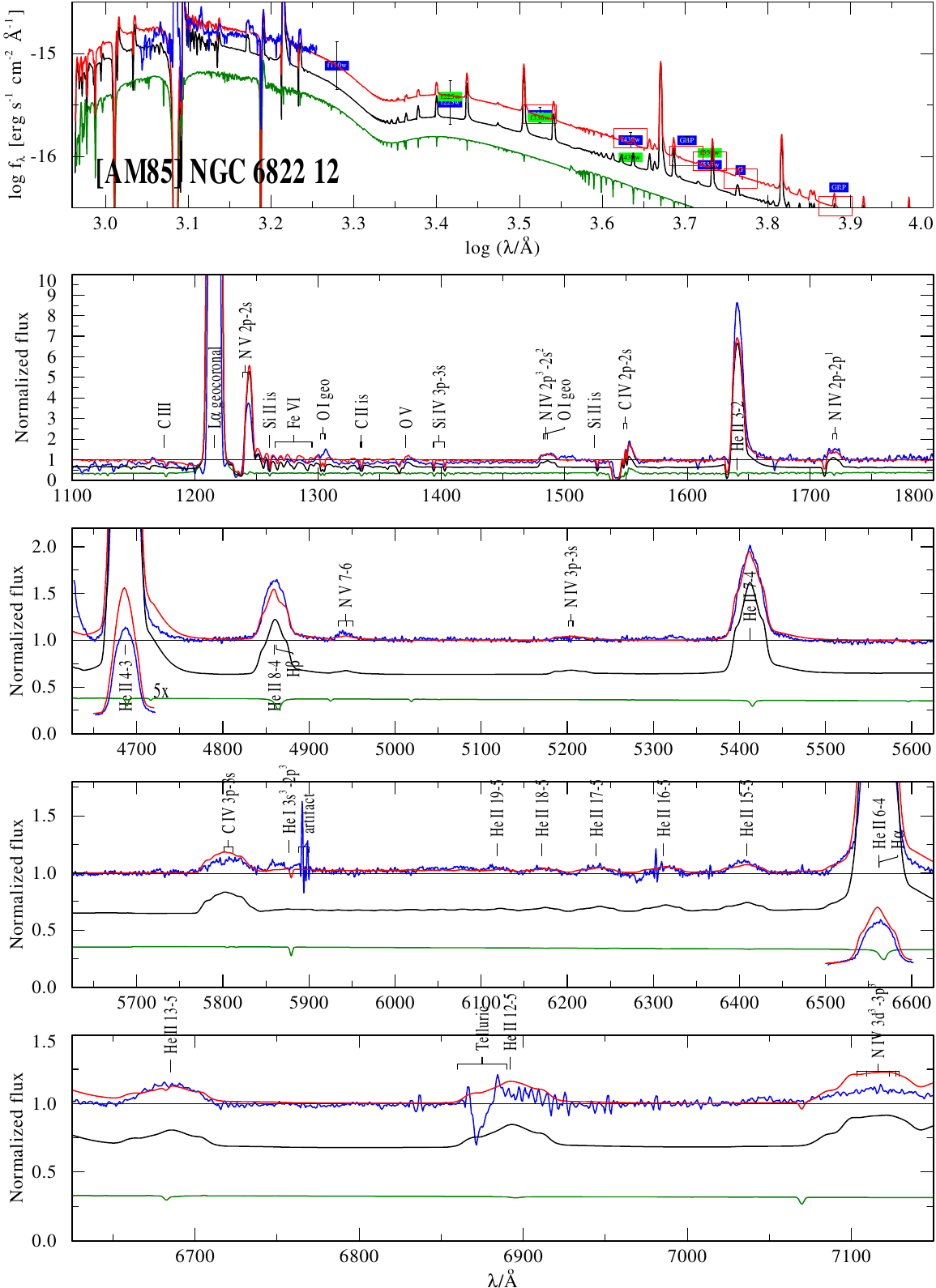}
	\end{center}
    \caption{Comparison between observed and synthetic spectra for star \#12. The observed spectrum is shown in blue. The composite model in red is the sum of the WR and O star models, shown by the black solid lines and green dot-dash lines respectively. In the upper panel, blue boxes indicate photometry extracted from HST images and from Gaia (see Sect.~\ref{Obs_Photometry}). For this star, the additional green boxes indicate photometry by \citet{NGCWR_Bianchi2001}. Red open boxes are centred on synthetic photometry, as calculated from the model flux and the corresponding filter function. The model parameters are given in Table~\ref{Stellar_parameters_all}.}
    \label{WR12_fullmasterplot}
\end{figure*}

The brightest star in our sample is \#12 (Fig.~\ref{WR12_fullmasterplot}). The first optical spectra were published by \citet{NGCWR_Bianchi2001a} as part of an analysis of massive stars in NGC~6822 (the star is named LB-f2-75 in their paper), and by \citet{WR12_Abbot2004}. 

We classify the star as WN3-4, in agreement with previous works. Several diagnostic lines  (N\,{\scshape iv} $\lambda$3479-3484\,\AA\,and $\lambda$4058\,\AA, N\,{\scshape v} $\lambda$4603\,\AA\,and $\lambda$4619\,\AA) are outside the range of our observed spectrum so it is difficult to determine the exact relative strengths of the N\,{\scshape iv} and N\,{\scshape v} lines. As such, we refrain from discriminating between WN3 and WN4 types.

\citet{NGCWR_Bianchi2001} estimated an effective temperature of 38~kK by fitting the observed photometric magnitudes to theoretical magnitudes based on the LTE models by \citet{WR12_Bessel1998}. Our analysis strongly favours a much hotter model. SMC WN-star PoWR grid models at temperatures $\sim$$40$\,kK predict P Cygni profiles for He\,{\scshape i} $\lambda$4626\,\AA\,and C\,{\scshape iv} $\lambda$7060\,\AA\,that are similar in strength to H$\alpha$ $\lambda$6560\,\AA. This is not the case in the observed spectra. We estimate the temperature as $T_{\ast}=85 \pm 10$~kK using the observed ratio between N\,{\scshape iv} and N\,{\scshape v} lines. A high temperature and luminosity for \#~12 has also been favoured by \citet{WR12_Abbot2004} who found $T_\ast = 100$~kK and $\log(L/L_\odot) = 6.11$, using the CMFGEN  model atmospheres \citep{CMFGEN_Hillier1998} code. However, the spectrum analysed by \citet{WR12_Abbot2004} did not include the UV range and covered a much narrower range in the optical compared to our data. In this work, we have access to several more diagnostic lines (e.g. N\,{\scshape iv} $\lambda$7100-7130\,\AA), likely explaining the slight difference in temperature estimates. Note that when \#~12 is modelled as a single star, similar to \citet{WR12_Abbot2004} we determine $\log(L/L_\odot) = 6.15$ from the SED fit (Fig.~\ref{WR12single_fullmasterplot}).

Our spectral modelling leads to an update of the metallicity. The CNO abundances have been kept at values typical for the SMC WN-type stars \citep{WRSMC_Shenar2016}. However, in order to reproduce the forest of iron lines in the UV spectrum, the abundance of iron and other heavy elements is decreased to $X_\text{Fe}$ = $1\times10^{-4}$, i.e.\ a third of the SMC value. 

We constrain the hydrogen content to less than 5\%. If no hydrogen is included in a model, the increase in the strength of He lines must be balanced by a reduction in mass-loss rate. This leads to poorer fits for the nitrogen and carbon lines, suggesting that the star does contain some small amount of hydrogen. 

\citet{NGCWR_Bianchi2001a} interpreted the emission bump at $\lambda$5850\,\AA\,as a He\,{\scshape i} line. In our observed spectrum, this feature appears to be strongly broadened with a dip in the centre. There are no other notable stellar lines at these wavelengths that could be producing a second peak. The dip is well-modelled as an absorption line from the O-star companion spectrum, but the emission feature is not reproduced by our models at $T_\ast \approx 85$~kK, potentially favouring a cooler temperature. However, this is not supported by the observed ratio of N\,{\scshape v}/ N\,{\scshape iv}. 

Further indication for binarity is the weak absorption feature in the UV spectrum which we tentatively interpret as a C\,{\scshape iii} $\lambda$1175\,\AA\ line. Considering together with the unusual He\,{\scshape i}$\lambda$5850\,\AA\, feature, we assume that these arise from a binary companion. It is difficult to constrain the parameters of the companion as it gives only a small contribution to the spectrum. We choose an O-star grid model with $T_\ast = 38$~kK, $\log(g)=3.6$~cm~s$^{-2}$ and adjust the radial velocity in order to reproduce the C\,{\scshape iii} $\lambda$1175\,\AA\,line and the small asymmetries visible at the peaks of e.g. the He\,{\scshape ii} $\lambda$4860\,\AA\,line, without hindering the rest of the fit. For comparison, the best-fit single WR-star model is shown in Fig.~\ref{WR12single_fullmasterplot}.

As can be appreciated from Fig.\,\ref{WR12_fullmasterplot}, most of the observed emission line profiles are unusually broad and rounded compared to other WR stars. Such round line profiles have previously been observed in some WNE type (comprising the subtypes WN2-6) stars in the Galaxy and LMC \citep[][]{ROT_Shenar2014,Chene2019} but never in a lower metallicity galaxy. \citet{ROT_Shenar2014} showed that round emission line profiles can be explained by rotational broadening if part of the stellar wind corotates with the star. Such a situation could arise, for instance, in magnetic stars. We investigated this possibility and find that even with nearly critical rotational velocity at the photospheric equator, some corotation of the wind would be needed to match the observed line shapes. To enforce wind corotation up to a radius of 1.5~$R_\ast$, a magnetic field of many kG is required. Magnetic fields of this strength have not yet been detected in classical WR stars \citep[e.g.][]{delaChevrotiere2014, MAGNETICWR_Hubrig2016}.

Alternatively, the round shape of the emission line profiles could be attributed to the specific wind velocity field \citep{VELO_Lefever2023}. We tested this by adopting in the PoWR model a velocity field based on predictions from hydrodynamical models. As can be seen in Fig.~\ref{WR12_fullmasterplot}, the line shapes are well reproduced in this case. Figure~\ref{WR12_NOVELOfullmasterplot} shows the best-fit model for the star adopting a $\beta$-law for the velocity field and no rotation - the line shapes are clearly less well reproduced.

At the very blue edge of our optical spectrum we notice an increase in flux, likely explained as part of the N\,{\scshape v} $\lambda\lambda$4604, 4620\,\AA\, doublet. This line is clearly visible in the spectrum displayed in \citet{WR12_Abbot2004}. Our best model reproduces the N\,{\scshape iv} $\lambda$1720\,\AA, He\,{\scshape ii} $\lambda$4860\,\AA, He\,{\scshape ii} $\lambda$5412\,\AA\ lines quite well, but predicts somewhat too strong lines of N\,{\scshape v} $\lambda$1240\,\AA, C\,{\scshape iv} $\lambda$1550\,\AA\, and $\lambda$5801-12\,\AA, N\,{\scshape iv} $\lambda$7100-7130\,\AA. Tightening N-abundance estimates would be helpful to better reproduce these lines. 

\subsection{NGC 6822 4}\label{WR4_analysis}

\begin{figure*}
	\begin{center}	
		\includegraphics[width=0.7\linewidth]{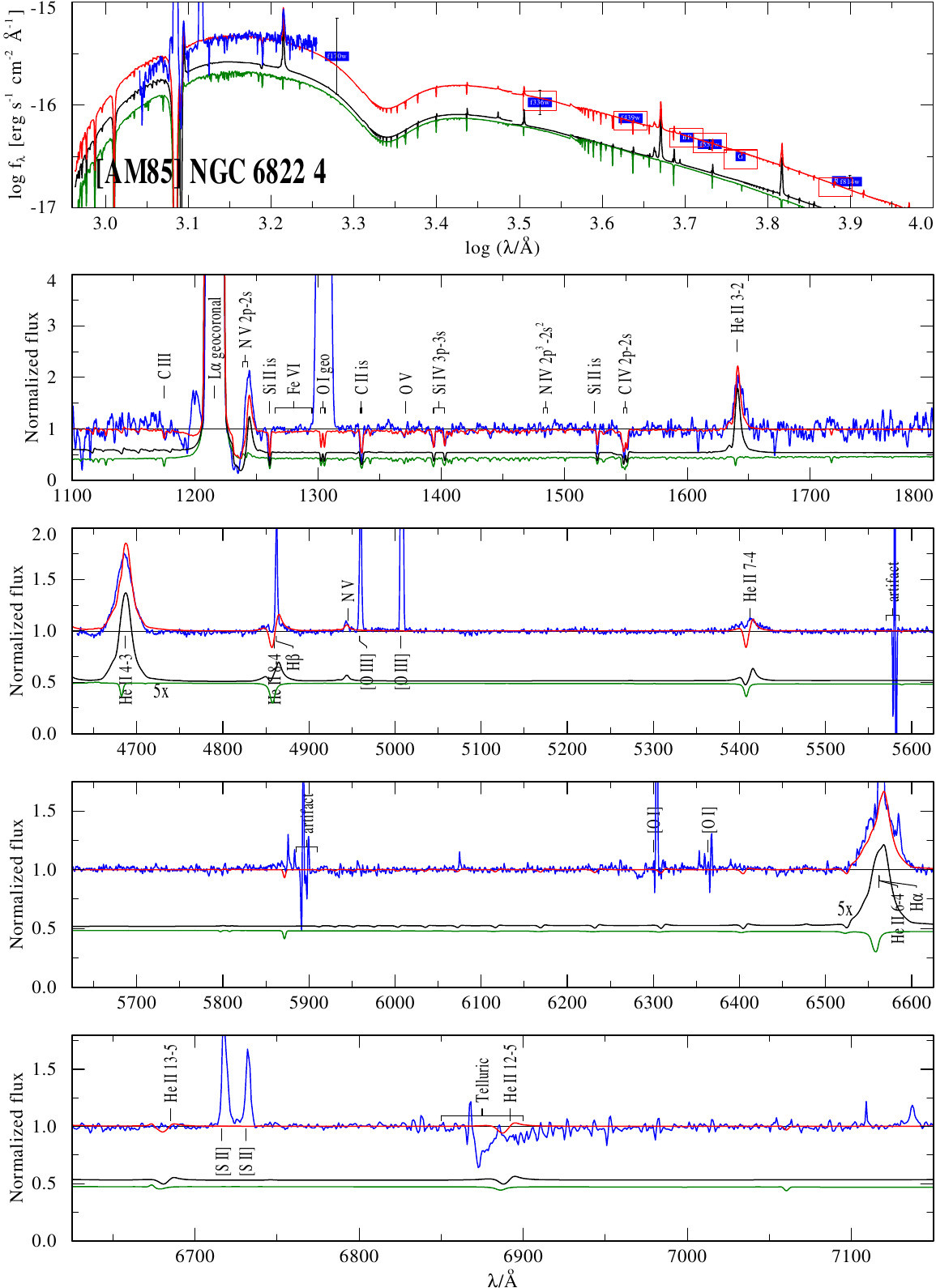}
	\end{center}
    \caption{Same as Fig.~\ref{WR12_fullmasterplot} but for star \#4.}
    \label{WR4_fullmasterplot}
\end{figure*}

The observed and model spectra of star \#4 are displayed in Fig.~\ref{WR4_fullmasterplot}. The star has a WN3 type, since the N\,{\scshape v} lines are pronounced and the N\,{\scshape iv} lines are absent. The line profiles are narrower and more triangular, i.e.\ more typical for WR stars, compared to the spectra of stars \#5 and \#12. The spectrum also features prominent nebular lines ([O\,{\scshape ii}] $\lambda$4958\,\AA\,\& $\lambda$5006\,\AA, [S\,{\scshape ii}] $\lambda$6716\,\AA\,\& $\lambda$6731\,\AA,  [O\,{\scshape i}] $\lambda$6300\,\AA\,\& $\lambda$6364\,\AA). The stellar lines H$\alpha$ and H$\beta$ are blended with narrow nebular emission. The hydrogen content is constrained to 40\% while other abundances have been kept at the SMC WN-star grid model values. 

The absence of N\,{\scshape iv} emission implies a  $T_\ast\gtrsim70$~kK though it is otherwise not well constrained ($T_\ast= 84 \pm 10$~kK). It is difficult to place a strict upper limit on the temperature, but we find that higher temperature models produce an overall poorer fit. 

The presence of C\,{\scshape iii} $\lambda$1175\,\AA\,and O\,{\scshape v} $\lambda$1371\,\AA\,absorption lines indicate that star \#4 is a binary with a hot O star companion. We choose an O star grid model with $T_\ast = 44$~kK, $\log(g)=4.2$~cm~s$^{-2}$ to reproduce these features and provide a good fit for the C\,{\scshape iv} $\lambda\lambda$ 1548, 1550\,\AA\,line. Furthermore, the N\,{\scshape v} $\lambda$1240\,\AA\ line has a P Cygni profile. Its absorption feature is partially filled, likely indicating a contribution from a companion star (alternatively, the desaturation of the N\,{\scshape v} P Cygni absorption could be due to an overlap with the Ly$\alpha$ geocoronal emission). In general, a single-star model produces somewhat poorer spectral fit -- our best single-star model is shown in Fig.~\ref{WR4single_fullmasterplot}.

\subsection{NGC 6822 5}\label{WR5_analysis}

\begin{figure*}
	\begin{center}	
		\includegraphics[width=0.7\linewidth]{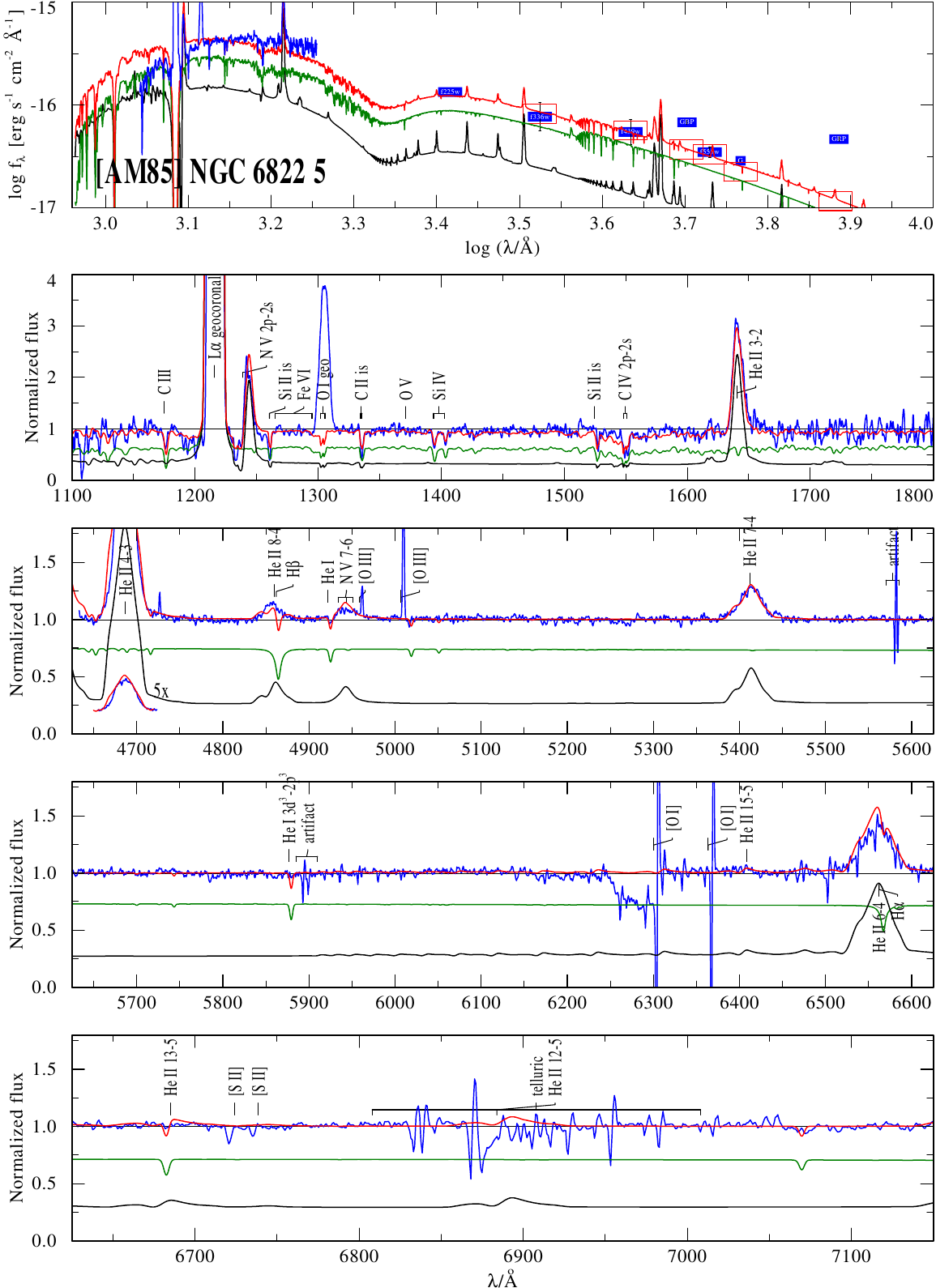}
	\end{center}
    \caption{Same as Fig.~\ref{WR12_fullmasterplot} but for star \#5.}
    \label{WR5_fullmasterplot}
\end{figure*}

The observed and best model spectra of \#5 are displayed in Fig.~\ref{WR5_fullmasterplot}. We classify \#5 as WN3 type since no N\,{\scshape iv} emission is observed. The star is hydrogen-free. The absence of N\,{\scshape iv} emission implies a lower limit of $T_\ast \gtrsim 80$~kK, but otherwise the temperature of this star is not well constrained. A temperature of $T_\ast=110 \pm 10$~kK is adopted to achieve, in particular, a good fit for the C\,{\scshape iv} $\lambda\lambda$ 1548, 1550\,\AA\,line.

Similar to star \#4, the absorption feature of the P Cygni profile of N\,{\scshape v} $\lambda$1240\,\AA\, is partially filled. As mentioned in Sect.~\ref{WR4_analysis}, this could either be due to an overlap with the Ly$\alpha$ geocoronal emission, or indicate a contribution from a companion star. A further indicator of binarity is the strong C\,{\scshape iii} $\lambda$1175\,\AA\, absorption line. The strength of this line indicates a cooler B-type companion. This is corroborated by the shape of the  C\,{\scshape iv} $\lambda\lambda$ 1548, 1550\,\AA\, line. Our tests show that the secondary temperature of 20\,--\,30\,kK allows us to match this line in the composite spectrum. We assume that the companion is rotating with $\varv \sin{i}= 300$\,km\,s$^{-1}$. The synthetic spectrum of the companion slightly over-predicts the strength of the H$\beta$ absorption line, suggesting a higher temperature. Future multi-epoch observations are required to clarify the nature of the companion star. For comparison the best single-star model is shown in Fig.~\ref{WR5single_fullmasterplot}.

\subsection{NGC 6822 3}\label{WR3_analysis}

\begin{figure*}
	\begin{center}	
		\includegraphics[width=0.7\linewidth]{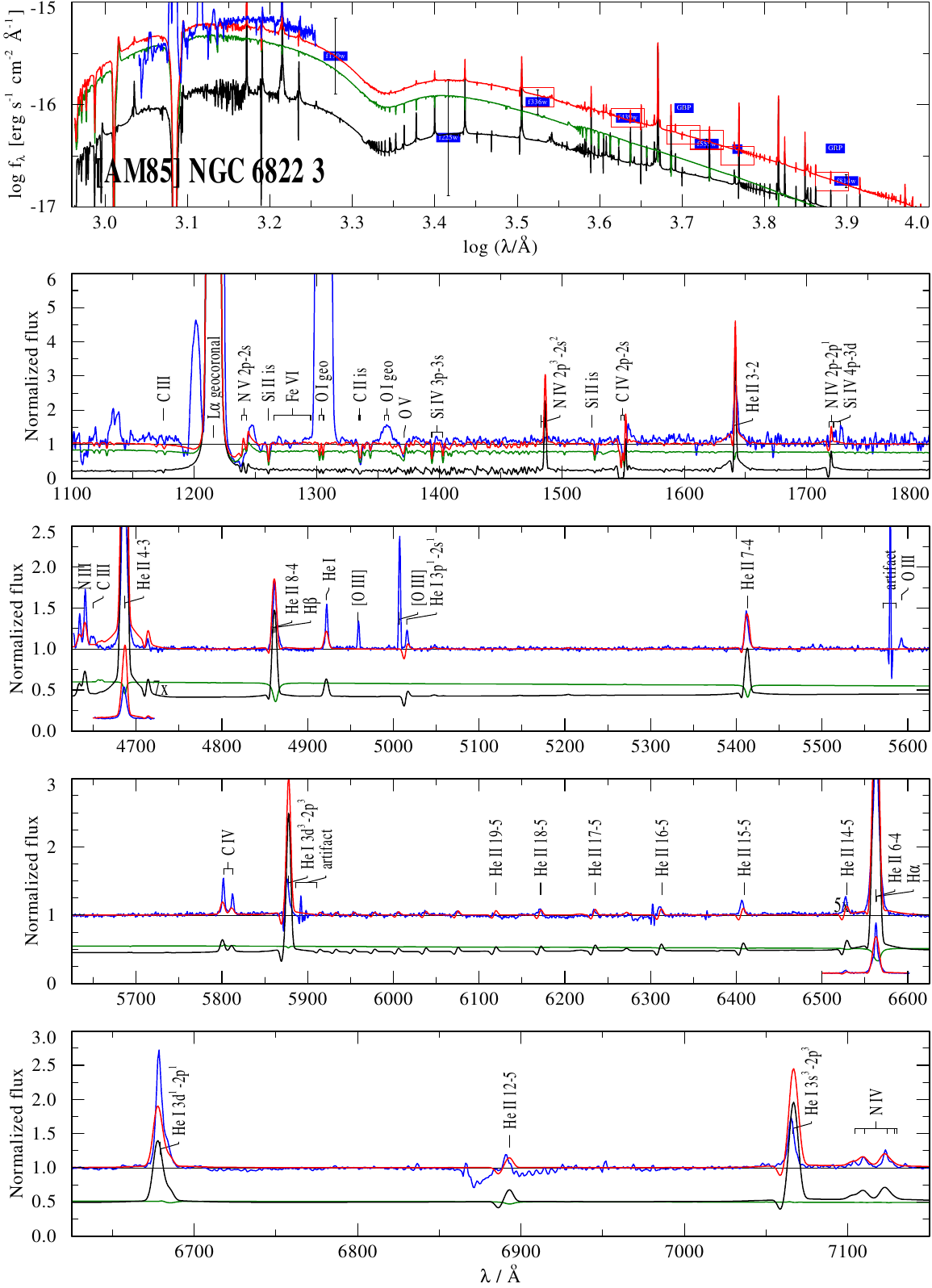}
	\end{center}
    \caption{Same as Fig.~\ref{WR12_fullmasterplot} but for star \#3.}
    \label{WR3_fullmasterplot}
\end{figure*}

\begin{figure}[]
	\begin{center}	
	\includegraphics[width=0.9\linewidth]{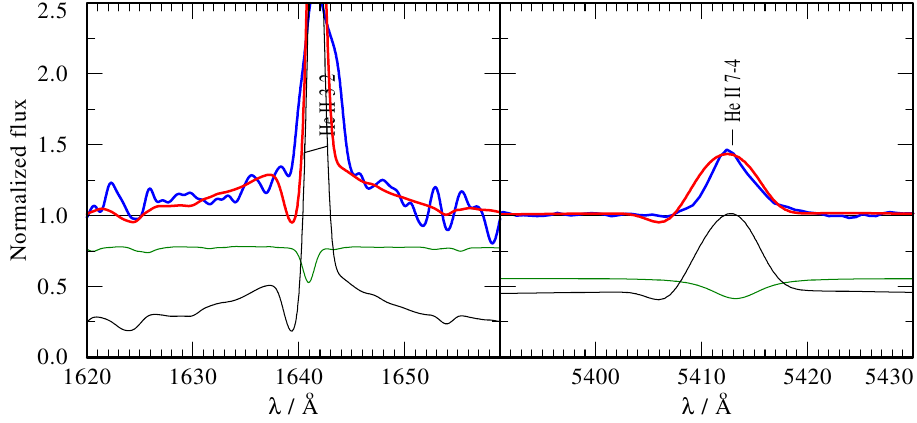}
	\end{center}
    \caption{Zoom into the spectral fit of star \#3. The observed spectrum is shown in blue. The composite model in red is the sum of the WR and O star models, shown by the black and green lines respectively. The He\,{\scshape ii} $\lambda$1640\,\AA\,line in UV seems to indicate a greater terminal wind velocity than the optical line He\,{\scshape ii} $\lambda$5412\,\AA\,(see text). The WR component shows different radial velocities in the UV and optical observations, which have been taken at different epochs (see Tables~\ref{Obs_optical} and \ref{tab:Obs_UV}). The UV and optical observations are corrected correspondingly.}
    \label{WR3_zoomplot_He}
\end{figure}

We classify star \#3 as WN6 -- the N\,{\scshape iv} lines are weak, and the N\,{\scshape iii} $\lambda\lambda$4634–4641\,\AA\ triplet is weaker than the He\,{\scshape ii} $\lambda$4686\,\AA\,line (Fig.~\ref{WR3_fullmasterplot}). Several of the stellar lines may be blended with nebular emission, but it is difficult to tell given the limited spectral resolution of our observations. The lines in the optical spectrum are unusually narrow, corresponding to a terminal wind velocity of only $\varv_\infty=200 \pm 100$ km~s$^{-1}$. However, the width of the UV lines suggest a somewhat larger wind speed (see Fig.~\ref{WR3_zoomplot_He}). This likely indicates that the velocity field deviates from a $\beta$-law. 

The spectrum of \#3 contains a range of He\,{\scshape i} and He\,{\scshape ii} emission lines, and it has been especially difficult to match the strengths of all the lines at once. From comparing the He\,{\scshape i} and He\,{\scshape ii} lines, and using the C\,{\scshape iv} $\lambda$5801-12\,\AA and N\,{\scshape iii} $\lambda$4640\,\AA\ lines, the temperature is estimated to be $T_\ast \approx 63$~kK. 

The observed He\,{\scshape i} $\lambda$4921\,\AA, and $\lambda$5016\,\AA, and He\,{\scshape ii} $\lambda$6560\,\AA\, lines may be contaminated by nebular emission. In this case, the modelled lines are slightly too strong. On the other hand, the model He\,{\scshape i} $\lambda$6678\,\AA\ line is slightly too weak. The N\,{\scshape iv} $\lambda$1490\,\AA, He\,{\scshape ii} $\lambda$1550\,\AA, and He\,{\scshape i} $\lambda$4686\,\AA\ lines are consistently too strong in all models where other lines fit.

The observed C\,{\scshape iii} $\lambda$4650\,\AA\ line is of comparable strength to the nearby N\,{\scshape iii} $\lambda\lambda$4634–4641\,\AA\,triplet. This is not reproduced by the model and strongly resembles the Of?p phenomenon observed in the spectra of magnetic stars \citep[][, discussed further in Sect.~\ref{WR3_evo}]{OFP_Walborn1972}.

There are indications that star \#3 is a binary. Firstly, there is a difference in the radial velocity by $\approx 200$~km s$^{-1}$ between the UV and optical observations, which have been taken at different epochs (see Tables~\ref{Obs_optical} and \ref{tab:Obs_UV}). Secondly, we could not achieve a good fit to the observed UV spectrum using a single-star model (see Fig.\ref{WR3single_fullmasterplot}). When the optical spectrum is reasonably well fit, the N\,{\scshape iv} $\lambda$1490\,\AA, C\,{\scshape iv} $\lambda$1550\,\AA, He\,{\scshape ii} $\lambda$1640, and N\,{\scshape iv} $\lambda$1718\,\AA\ lines are far too strong and the absorption part of the C\,{\scshape iv} $\lambda$1550\,\AA\ P Cygni profile is fully saturated. 

\begin{figure}[H]
	\begin{center}	
		\includegraphics[width=0.9\linewidth]{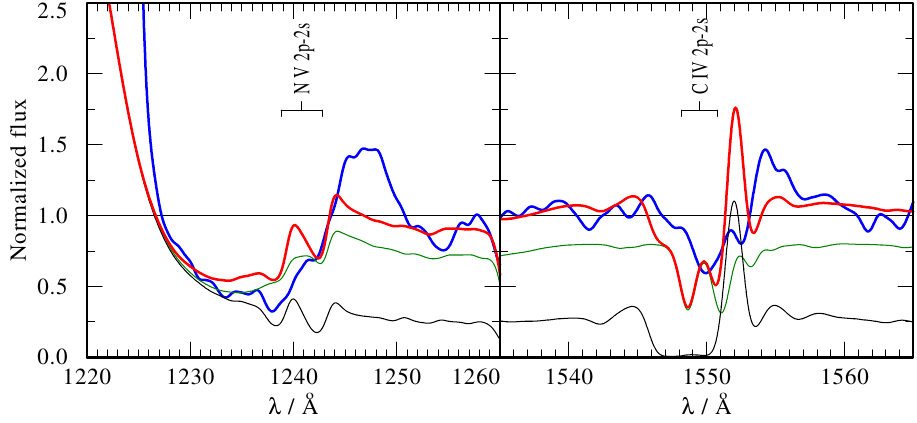}
	\end{center}
    \caption{N\,{\scshape v} $\lambda$1240\,\AA\,and C\,{\scshape iv} $\lambda$1550\,\AA\,lines for star \#3. The composite model in red is the sum of the WR and O star models, shown by the black and green lines respectively. The relative offsets of the model continua correspond to the light ratio between the two stars.}
    \label{WR3_zoomplot_NV_CIV}
\end{figure}

Additionally, the observed N\,{\scshape v} $\lambda$1240\,\AA\ P\,Cygni profile is broad and cannot be reproduced as originating in the WR star wind, with the terminal wind velocity as inferred from fitting optical lines. We thus attribute this line to a more luminous hot O star companion. An O-star synthetic spectrum is selected from the OB star model grid to reproduce the O\,{\scshape v} $\lambda$1371\,\AA\, and C\,{\scshape iii} $\lambda$1175\,\AA\, lines. We include an X-ray radiation field in the O-star model, however this has little impact on the emergent spectra. The luminosities and mass-loss rates are adjusted to fit the observed  N\,{\scshape v} $\lambda$1240\,\AA\ and C\,{\scshape iv} $\lambda$1550\,\AA\ lines (Fig.~\ref{WR3_zoomplot_NV_CIV}). 

\begingroup
\setlength{\tabcolsep}{4.5pt} 
\begin{table*}\centering
	\caption{Derived stellar and wind parameters of all four known WR stars in the NGC~6822 galaxy}
		\begin{tabular}{c|cc|cc|cc|cc}
			\hline\hline
             & \multicolumn{2}{|c|}{Star \#3} & \multicolumn{2}{|c|}{Star \#4} & \multicolumn{2}{|c|}{Star \#5} & \multicolumn{2}{|c}{Star \#12} \\

			Properties/Stars & WR & Companion & WR & Companion& WR & Companion & WR & Companion \\
            \hline
			
			$E(B-V)$ [mag]	 & \multicolumn{2}{|c|}{0.28$\pm0.02$} & \multicolumn{2}{|c|}{0.35$\pm0.02$}  & \multicolumn{2}{|c|}{0.25$\pm0.02$} & \multicolumn{2}{|c}{0.24$\pm0.02$}\rule[0mm]{0mm}{4mm}\\
			
			$T_\ast$ [~kK] & 63$\pm10$ & 50$\pm5$ & 84$\pm10$ & 44$\pm5$ & 110$\pm10$ & 27$\pm5$ & 85$\pm10$ & 38$\pm5$ \\

            $T_{2/3}$ [~kK] & 34$\pm10$ & 50$\pm5$ & 82$\pm10$ & 44$\pm5$ & 108$\pm10$ & 27$\pm5$ & 75$\pm10$ & 38$\pm5$ \\

            $\log(L)$ [$L_\odot$] & $4.5^{+0.3}_{-0.4}$ & 5.1$\pm0.3$ & 5.65$\pm0.3$ & 4.9$\pm0.3$ & 5.2$^{+0.5}_{-0.3}$ & 4.4$\pm0.3$ & 5.85$\pm0.3$ & 4.8$\pm0.3$\\
			
			$\log{R_\text{t}}~[R_\odot]$  & 0.38$\pm0.1$ & 2.58 & 1.22$\pm0.05$ & 2.95 & 0.64$\pm0.05$ & 4.0 & 0.73$\pm0.05$ & 2.52 \\
			
			$\log(\dot{M})$  [$M_\odot$yr$^{-1}$] & -5.7$\pm0.2$ & -7.2 & -5.55$\pm0.2$ & -7.8 & -5.4$\pm0.2$ & -9.3 & -4.7$\pm0.2$ & -7.2 \\

            $\log(\dot{M_t})$ & -3.58 & -6.58 & -5.22 & -6.97 & -4.60 & -7.91 & -4.49 & -6.137 \\

            $\varv_{\text{rad}}$ [km s$^{-1}$]	& -100$\pm$20$^b$ & -100$\pm$50 & -100$\pm$50 & -50$\pm$50 & -200$\pm$50 & -50$\pm$20 & -50 $\pm$20 & +200$\pm$50 \\

            $\varv_{\infty}$ [km s$^{-1}$]	& 200$\pm$50 & 3000 & 1700$\pm$100 & 1783 & 1600$\pm$100 & 2000 & 1600$\pm$100 & 2200 \\
            
			$M_V$ [mag]& -4.84$\pm0.3$ & ... & -4.73$\pm0.3$ & ... & -5.21$\pm0.3$ & ... & -5.32$\pm0.3$ & ... \\

			$X_\text{H}$ & $0.15\pm0.05$ & 0.7375 & $0.4\pm0.05$ & 0.7375 & $0^{+0.05}$ & 0.7375 & 0.05$\pm0.05$& 0.7375 \\

            $X_{\text{Fe}}/10^{-4}$ & 3 & 3.52 & 3 & 3.52 & 3  & 3.52 & 1$\pm0.5$ & 3.52 \\
            
            $X_\text{C}/10^{-5}$ & 2.5$\pm2$ & 21.4 & 2.5$\pm2$ & 21.4 & 2.5$\pm1$ & 21.4 & 2.5$\pm0.5$ & 21.4 \\
            
            $X_\text{N}/10^{-3}$ & 1.5$\pm0.2$ & 0.0326 & 1.5$\pm0.5$ & 0.0326 & 1.5$\pm0.2$ & 0.0326 & 1.5$\pm0.2$ & 0.0326 \\

            $X_\text{O}/10^{-5}$ & 2.5 & 113 & 2.5 & 113 & 2.5 & 113 & 2.5 & 113 \\

			$D_\infty$ & 4 & 10 & 4 & 10 & 4 & 10 & 4 & 10\\
            
			$R$ [$R_\odot$] & 1.5$^{+0.88}_{-0.66}$ & 5.3 & 3.2$^{+1.8}_{-1.1}$ & 4.9 & 1.1$^{+0.85}_{-0.39}$ & 8.5 & 3.9$^{+2.1}_{-1.3}$ & 5.8 \\

            $M~[M_\odot]^a$ & 4.6$^{+4.6}_{-2.8}$ & 20.6 & 19.2$^{+19.1}_{-9.6}$ & 13.7 & 10.6$^{+16.0}_{-6.4}$ & 7.3 & 25.5$^{+25.4}_{-12.7}$ & 19.5 \\

            $\log{Q_\text{H}}$~[s\textsuperscript{-1}] & 48.3 & 49.1 & 49.5 & 48.7 & 49.0 & 46.9 & 47.7 & 48.4 \\
            
            $\log{Q_\text{He\,\scshape I}}$~[s\textsuperscript{-1}] & 47.9 & 48.7 & 49.3 & 48.3 & 48.9 & 44.0 & 49.5 & 47.4 \\
            
            $\log{Q_\text{He\,\scshape II}}$~[s\textsuperscript{-1}] & 38.6 & 44.6 & 47.3 & 44.0 & 47.1 & 36.05 & ...$^c$ & 42.3 \\

			\hline
            
		\end{tabular}			
    \tablefoot{$^\text{a}$ estimated using the mass-luminosity relations for H-rich and H-free stars from \citet{MASSLUM_Grafener2011}. $^\text{b}$ from the optical lines; the UV spectrum indicates a different radial velocity shift of 320~km s$^{-1}$. $^\text{c}$ optically thick at outer boundary for He\,{\scshape ii} ionizing photons.}
	\label{Stellar_parameters_all}
\end{table*}
\endgroup

\section{Discussion}

\subsection{On the origin and evolution of WR stars in NGC\,6822}\label{evo_tracks}

\begin{figure}
	\begin{center}
		\includegraphics[width=\linewidth]{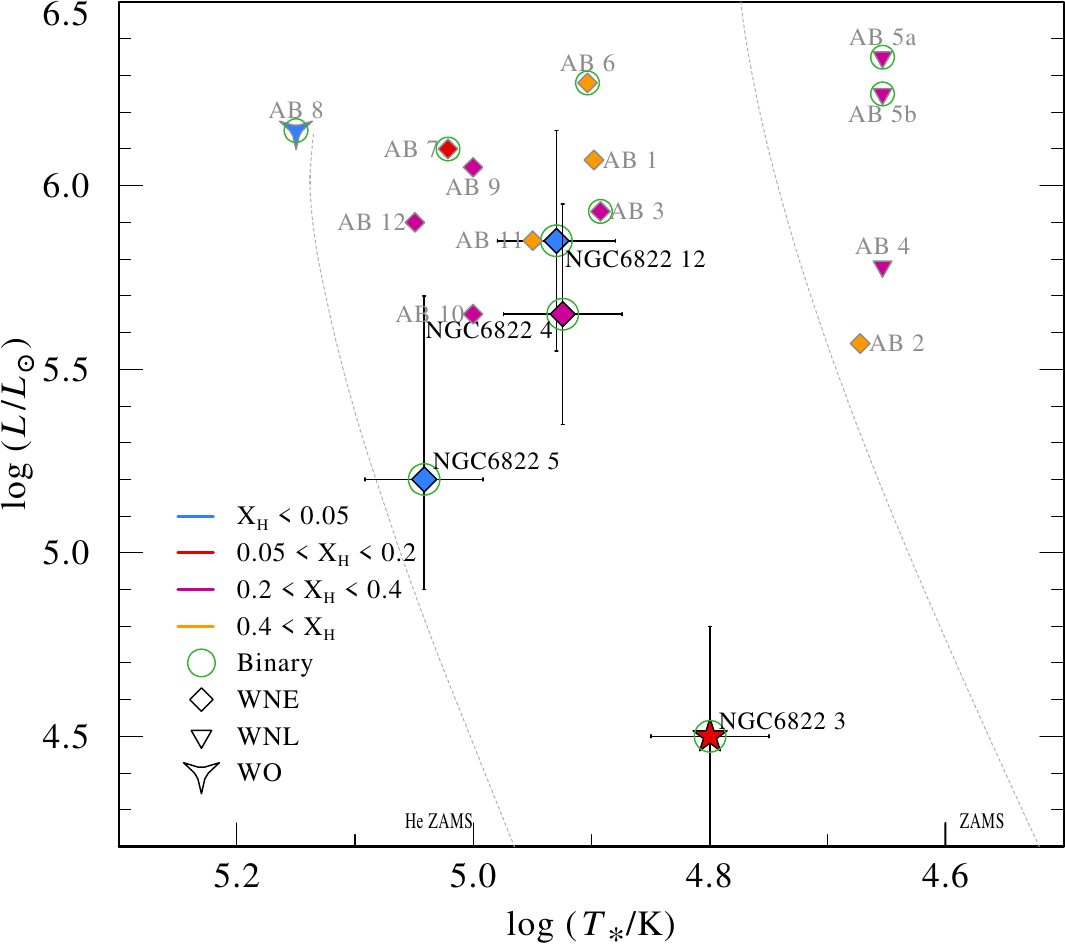}
	\end{center}
    \caption{HRD of the WR stars in NGC~6822 (larger symbols with error bars) and in the SMC \citep[smaller symbols without error bars][]{WRSMC_Hainich2015,WRSMC_Shenar2016}. The encircled symbols indicate binaries. Hydrogen content is colour coded,  the meaning of colours and symbols is explained in the legend. }
    \label{hrd+smc}
\end{figure}

Fundamental stellar parameters derived from detailed spectroscopic analysis allow us to place the sample stars on the HRD (Fig.\,\ref{hrd+smc}). Star \#3 has very unusual properties and is discussed separately in Sect.~\ref{WR3_evo}. 

\begin{figure}
	\begin{center}	
		\includegraphics[width=\linewidth]{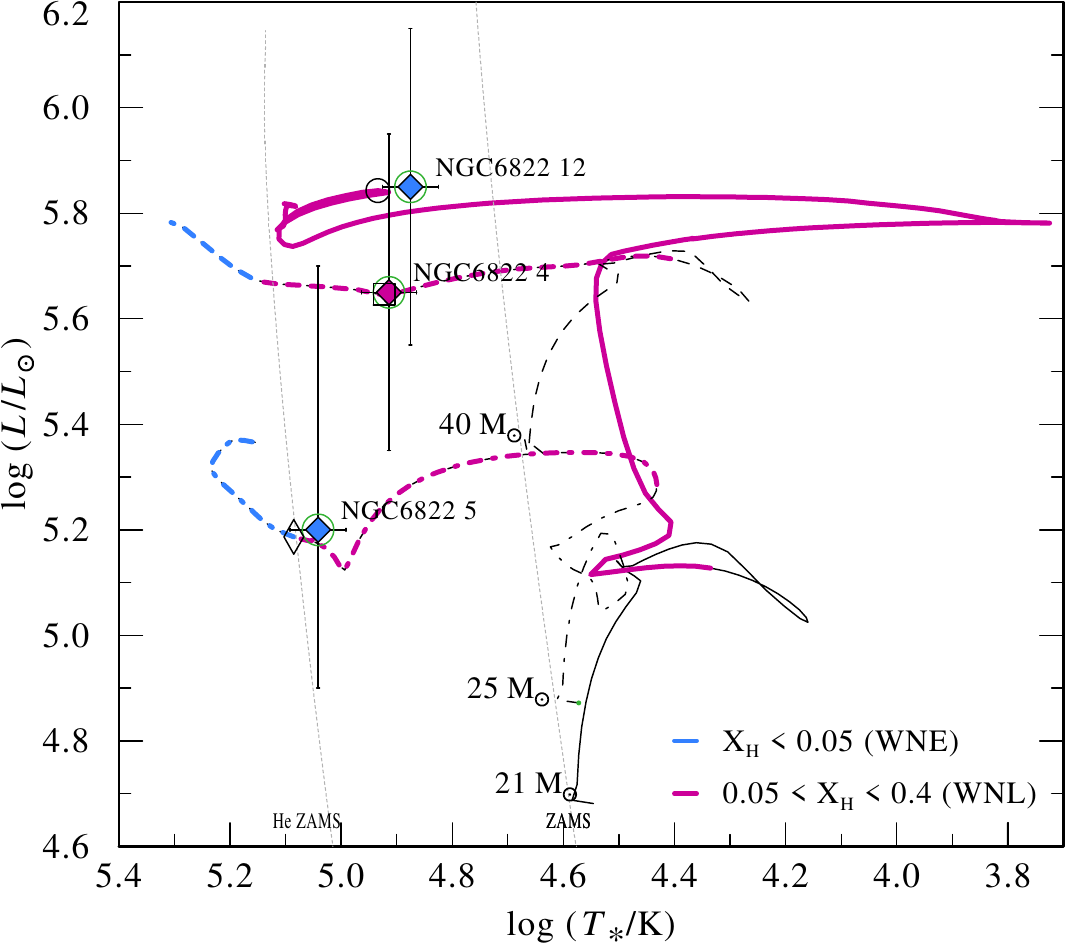}
	\end{center}
    \caption{
    HRD showing the positions of WR stars in NGC\,6822 according to the spectroscopic analysis (Sect.~\ref{Stellar Atmosphere Modelling}), and the best fitting binary BPASS evolutionary tracks labeled according to the initial mass of the primary. The solid line is for $M_{i, 1}=21\,M_\odot$, the dash-dotted line is for  $M_{i, 1}=25\,M_\odot$, and the dashed line is for $M_{i, 1}=40\,M_\odot$. The hydrogen content is colour coded, see legend. The secondary tracks are not shown for clarity.}
    \label{hrd+tracks}
\end{figure}

The location of the stars \#4, \#5, and \#12 on the HRD are compared with binary evolution tracks computed with the BPASS v.\,3 code\footnote{\url{bpass.auckland.ac.nz}}\citep[][]{Eldridge2017, Stanway&Eldridge2018}. Each track is calculated for  the metallicity  $Z = 0.002$ and is defined by the initial primary mass $M_{i, 1}$, orbital period $P_i$, and mass ratio $q_i$ = $M_{i,2}$/$M_{i,1}$. The tracks are calculated at intervals of 0.2 on 0 < $\log P$[d] < 4, and 0.2 on 0 < $q_i$ < 0.9. The mass range of 10 < $M_{i,1}$ < 150 $M_\odot$ is considered; the tracks are calculated with steps of 10 -- 30 $M_\odot$. We use a $\chi^2$ fitting procedure to find a track at age $t$ which best reproduces the observables $\log T_{\ast, 1}$, $\log T_{\ast, 2}$, $\log L_{1}$, and $\log L_{2}$.  For all three stars, $\chi^2\sim 1$, and the best-fit evolutionary tracks are within 1$\sigma$ of almost all stellar parameters. The results are shown in Fig.~\ref{hrd+tracks} and Table~\ref{tab:Model_parameters}.

Comparisons of the empirically determined positions of the sample stars on the HRD with binary evolutionary tracks suggest that stars \#4, \#5, and \#12 underwent a Roche lobe overflow (RLOF) phase before the primary reached the WR phase. 

The position of star \#4 on the HRD is matched by the track with $M_i = 40$~M$_\odot$. The RLOF phase initiates at an age of 5.0 Myr and lasts $\approx 0.1$\,Myr. During this time, the primary's mass decreases from 38.6 to 18.6~M$_\odot$. Almost all of the hydrogen is stripped during the RLOF phase, at the end of which the hydrogen mass fraction has decreased to $X_\text{H}$ = 0.27. We note that the measured $X_\text{H}$ = 0.4 and that the position of star \#4 on the HRD can be similarly well-reproduced by single-star tracks (see Appendix \ref{app:evo}). If this star is in a wider binary and avoided RLOF, then the hydrogen fraction would be closer to the observed value.

According to the evolutionary model, star \#5 started its life with $M_i = 25~$M$_\odot$. The mass transfer phase  was initiated after 6.7\,Myr and lasted $\approx 1$\,Myr. During this period, the mass of the primary decreased by $\approx 15\,M_\odot$ and the hydrogen fraction dropped to $X_\text{H}$ = 0.2. 

The position of the star \#12 on the HRD is best reproduced by an evolutionary track with $M_i = 21~$M$_\odot$. According to the BPASS tracks, the WR star has a current mass of only 8~M$_\odot$. This is significantly below  $M\approx 26\,M_\odot$ inferred from the mass-luminosity relation (Sect.\,\ref{WR12_analysis}). Taking the evolutionary mass at a face value, this huge discrepancy could be explained if the star is on very advanced evolutionary stage at the end of the He-burning phase. In this case, according to theoretical predictions by \citet{MINLUM_Sander2020}, the huge $L/M$-ratio should trigger a very strong mass-loss rate $\log{\dot{M}}>-4$, which is ruled out by observations. A more feasible current mass of 19~M$_\odot$ is obtained from a track with an initial mass of 50~M$_\odot$. This track crosses the location of star \#12 on an HRD, however the track model is only hydrogen free at temperatures T$_\ast \gtrsim 125$~kK, which is clearly incompatible with observed parameters. Reducing uncertainty on luminosity and further constraining the binary parameters should help with resolving this conundrum. 

We note that the best fit track model has a slightly higher hydrogen fraction, X$_\text{H}$ = 0.1, than the empirically derived X$_\text{H}$ = 0.05 $\pm$ 0.05 (and therefore misses the WNE colour-coding in Fig.~\ref{hrd+tracks}). The mass transfer phase initiates after 8.9\,Myr and lasts $< 0.1$\,Myr. The mass of the primary decreases by $\approx 13\,M_\odot$ and the hydrogen fraction drops to $X_\text{H}$ = 0.3 during this period.

The uncertainties on the luminosities of WR stars are quite large, driven primarily by the uncertainty of the light ratio of components in binaries. However, the comparison with evolutionary models suggests that all three WR stars are have been stripped of hydrogen during RLOF in binary systems.
 
\subsection{Comparison between properties of the WR star population in the SMC and NGC~6822 galaxies}

The similar metallicities of NGC~6822 and the SMC is also confirmed by our spectral analyses -- except for star \#12, all WR stars have SMC-like CNO and iron abundances. Star \#12 has a lower iron content, implying that there are some chemical gradients across NGC\,6822.

The positions of the WR stars on the HRD are compared in Figs.~\ref{hrd+smc} and \ref{hrd+smc_t23}. It is clear that the WR stars in NGC~6822 differ from their cousins in the SMC.

\begin{figure}
	\begin{center}
		\includegraphics[width=\linewidth]{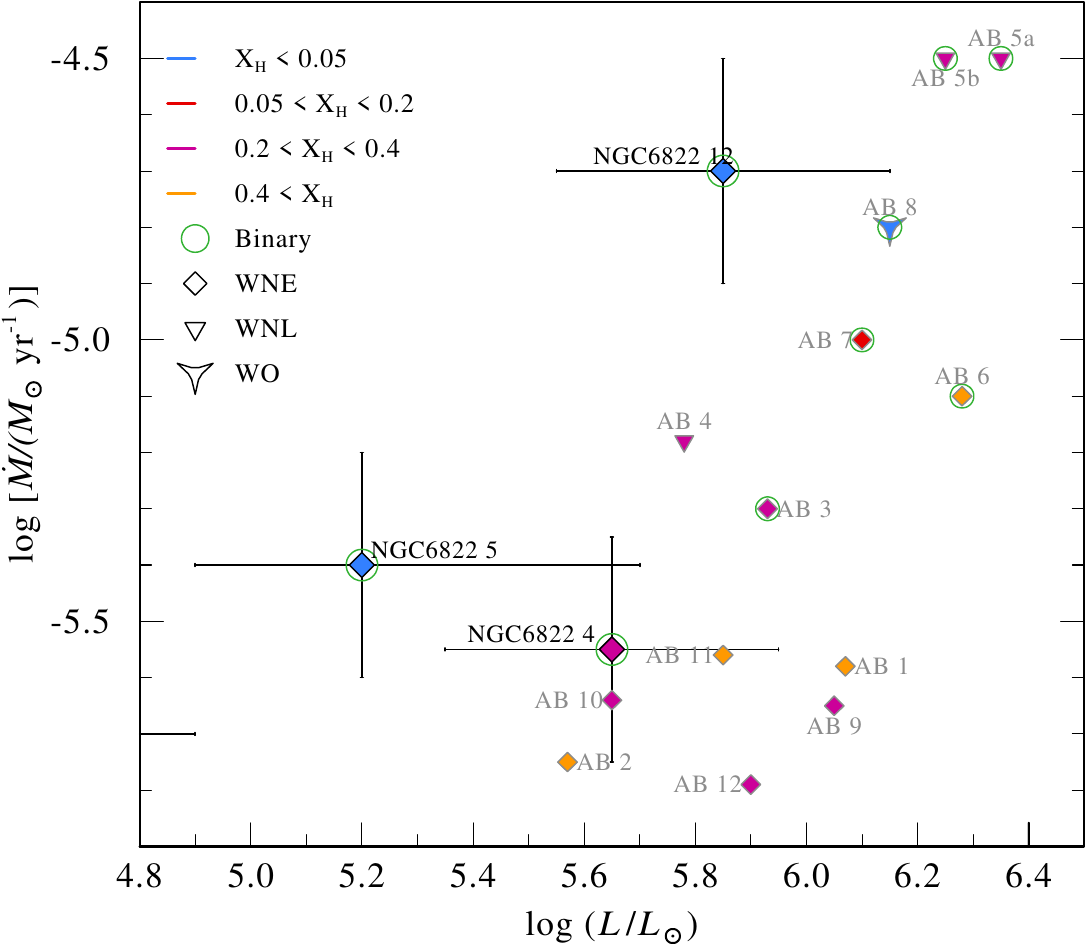}
	\end{center}
    \caption{Positions of the single and binary WR stars from NGC~6822 and the SMC on the $\log L$ - $\log \dot{M}$ plane. The meaning of the colours and symbols are indicated as in Fig. \ref{hrd+smc};}
    \label{mdot_vs_logl}
\end{figure}

It appears that WR stars in NGC\,6822 have modest masses but rather strong stellar winds (Table\,\ref{Stellar_parameters_all}, Fig.\,\ref{mdot_vs_logl}). This is not too surprising given that these low-$Z$ stars are likely close to the Eddington limit, and are capable of driving strong winds. Within the uncertainties on luminosity, the measured mass-loss rates are in agreement with recent theoretical and empiric recipes  \citep{MINLUM_Sander2020,WR_Shenar2020a, pauli+2025}. 

In the SMC, all WN stars contain significant amount of hydrogen. Prior to our work, it was not clear whether this is a fundamental property of stellar evolution in low-metallicity galaxies -- our results show that this is not the case. None of the SMC WR stars are similar to \#12 and \#5 in NGC\,6822. Although the luminosity and temperature of star \#12 are similar to the SMC stars, it has only a trace amount of hydrogen. Star \#5 is also hydrogen free, but  has luminosity lower than any WR star in the SMC. On the other hand, in NGC~6822, only star \#4 contains 40\%\ hydrogen and occupies a region in the HRD populated by the WR stars in the SMC. 

The samples of WR stars in these galaxies are too small to permit robust statistical comparisons. Nevertheless, the SMC hosts approximately three times as many WR stars as NGC~6822. If metallicity were the primary factor governing the formation and evolution of WR stars, then these two galaxies would exhibit similar distributions of WR star properties. Consequently, the smaller sample in NGC~6822 should be consistent with being drawn from the SMC population. However, this is not the case. 

The contrasting H-free WN populations of NGC~6822 and the SMC cannot be readily explained by metallicity alone. We therefore interpret this difference as a consequence of distinct star-formation histories in these galaxies. This interpretation supports the finding of \citet{TRACKS_pauli2026}, who showed that, under the assumption of a continuous star-formation history, some luminous H-free WN stars would be present in the SMC.

\subsection{Is [AM85]~NGC~6822 3 a strongly magnetic quasi-WR star? } \label{WR3_evo}

Star \#3 is enigmatic. It formally has a WN\,6 spectral type but a very slow stellar wind (Table\,\ref{Stellar_parameters_all}). On the HRD, it is located in the area occupied by the stripped helium stars stars with $M_{\rm i}\sim 10\,M_\odot$ \citep{Yungelson2024}. Population synthesis predicts that stripped He-stars are abundant in galaxies \citep[e.g.][]{Dionne2006,Hovis2025}. However, only a few such objects have been identified so far, with some of them having incomplete stripping \citep[e.g.][]{STRIPPED_DroutGotberg2023, STRIPPED_Ramachandran2024}. Among these sources, none has a WR-type spectrum.

The smoking gun pointing to a different nature of star \#3 is the notable presence of the emission line complex around C\,{\scshape iii} $\lambda$4650\,\AA. This complex is not observed in typical WN stars and is not reproduced by our atmosphere model. It is a defining diagnostic of the Of?p spectral type characteristic for strongly magnetic O stars \citep{Walborn2010}. The baseline CNO abundances are significantly lower in NGC\,6822 compared to the Galaxy, hence it is understandable that the strengths of the Of?p line complex is somewhat reduced.

The observed spectrum of star \#3 closely resembles the strongly magnetic merger product in the HD~45166 binary \citep[][ and references therein]{HD45166_Shenar2023}. In addition to spectroscopic similarities, both stars \#3 and the quasi-WR (qWR) component in the HD~45166 system occupy a similar area in the HRD. Star \#3 is flagged in the Gaia DR3 catalogue as an eclipsing binary with a possible period of 13.4\,d \citep{WR3_Gaia2022}; however, this requires further confirmation (see Sect.\ref{WR3_analysis}).

Hence, taking into consideration the spectral type of the star \#3, the presence of the Of?p line complex, the low inferred wind speed of $\sim 200$\,km\,s$^{-1}$, and its location in the HRD  we suggest this object is a magnetic qWR star like the one in HD~45166.

It is highly surprising to discover a second object of this type among the very small population of WR stars in the NGC~6822 galaxy. On the other hand, if our conjecture is correct and this a magnetic star, its presence in a low-metallicity galaxy is not fully unexpected. It has been shown that it is possible to form such stars in a failed common envelope merger scenario \citep{HD45166_Picco2026} -- the high magnetic field strength is due to amplification during the merger event, and likely does not depend on metallicity. Furthermore, the WN-type emission line spectrum arises from material trapped in the magnetosphere and hence spectral appearance depends on metallicity only weakly.  

\section{Summary and conclusions}

This work presents the first quantitative spectroscopic analysis of all four WR stars known in the galaxy NGC~6822 (Table\,\ref{tab:Star_names}). The galaxy has a low metallicity $Z\sim 0.2Z_\odot$, similar to the SMC. New UV and optical spectra of each star are analysed by means of the non-LTE model atmosphere code PoWR. We discuss the evolutionary channels leading to the formation of WR stars in NGC~6822 and compare the WR star population in NGC~6822 and the SMC. Our key conclusions are as follows:

\begin{itemize} 
    \item All four stars are spectroscopically confirmed as WN-type. We detect signatures of companion stars in each WR star spectrum, therefore each object in our sample is modelled as a binary. The derived stellar and wind parameters are provided in Table~\ref{Stellar_parameters_all}. Alternative single star solutions are presented in Appendices\,\ref{App:models} and \ref{app:evo}.\\

    \item We report the first detection of hydrogen-free WN-type stars at SMC-like metallicities -- stars \#5 and \#12 are essentially hydrogen free ($X_\text{H} \leq 0.05$). Star \#12 has iron abundances below the SMC baseline.  \\ 

    \item Spectroscopically derived stellar parameters are used to compare the locations of our sample stars on the HRD with BPASS binary evolutionary tracks. The results suggest that WR stars in NGC\,6822 stem from progenitors with initial masses in the range 20 -- 40 M$_\odot$. Significantly more massive progenitors are required by  single star evolutionary tracks.\\

    \item We present the discovery of a peculiar WR star. Star \#3 displays a spectrum characterised by comparatively narrow emission lines and an Of?p-type line complex, which highly resemble the spectrum of the magnetic merger product in the HD~45166 binary system. Future observations are needed to clarify the nature of the star \#3.\\

    \item The population of WR stars in NGC 6822 is distinct from the WRs in the SMC. This suggests that other factors beside metallicity may play a critical role in determining the properties of WR star population, such as recent star-formation history in the host galaxy. 
    
\end{itemize}

\noindent This first spectroscopic study of the complete known WR star population in the low-metallicity galaxy NGC~6822 provides a key benchmark for testing our understanding of the formation, observable properties, and final evolutionary stages of massive stars in metal-poor environments. Our results reveal a remarkable diversity in the physical and spectroscopic properties of WR stars at low metallicity, challenging current evolutionary predictions and highlighting the need for renewed observational and theoretical investigations to establish a more comprehensive picture of massive-star evolution under these conditions.

\begin{acknowledgements}
This paper is dedicated to the memory of Dr.\,Derck Massa who made profound contributions 
to the development of spectroscopic UV diagnostics of stellar winds.

R.Trigg acknowledges financial support by the Deutsches Zentrum f\"ur Luft-und Raumfahrt (DLR) grant FKZ 50 OR 2511.
AACS and RRL are supported by the Deutsche Forschungsgemeinschaft (DFG, German Research Foundation) in the form of an Emmy Noether Research Group – Project-ID 445674056 (SA4064/1-1, PI Sander). AACS and RRL further acknowledge support by the Baden-Württemberg Ministry of Science as part of the Excellence Strategy of the German Federal and State Governments.
VR acknowledges financial support by the Federal Ministry of Research, Technology and Space (BMFTR) via the Deutsches Zentrum f\"ur Luft- und Raumfahrt (DLR) grant 50 OR 2509 (PI Sander). This project was co-funded by the European Union (Project 101183150 - OCEANS).

TS acknowledges support from the Israel Science Foundation (ISF) under grant number 0603225041 and from the European Research Council (ERC) under the European Union's Horizon 2020 research and innovation program (grant agreement 101164755/METAL)

The authors acknowledge the important contribution to this work by Julian St\"ahle, who performed the initial reduction and analysis of optical observations when preparing his Master thesis.
 
\end{acknowledgements}

\bibliographystyle{aa}
\bibliography{refs}

@ARTICLE{Li+2024,
       author = {{Li}, Zhuowen and {Zhu}, Chunhua and {L{\"u}}, Guoliang and {Li}, Lin and {Liu}, Helei and {Guo}, Sufen and {Yu}, Jinlong and {Lu}, Xizhen},
        title = "{The Population Synthesis of Wolf─Rayet Stars Involving Binary Merger Channels}",
      journal = {\apj},
         year = 2024,
        month = jul,
       volume = {969},
       number = {2},
          eid = {160},
        pages = {160},
          doi = {10.3847/1538-4357/ad4da8},
archivePrefix = {arXiv},
       eprint = {2405.11571},
 primaryClass = {astro-ph.SR},
       adsurl = {https://ui.adsabs.harvard.edu/abs/2024ApJ...969..160L}
}

@ARTICLE{Walborn2010,
       author = {{Walborn}, Nolan R. and {Sota}, Alfredo and {Ma{\'\i}z Apell{\'a}niz}, Jes{\'u}s and {Alfaro}, Emilio J. and {Morrell}, Nidia I. and {Barb{\'a}}, Rodolfo H. and {Arias}, Julia I. and {Gamen}, Roberto C.},
        title = "{Early Results from the Galactic O-Star Spectroscopic Survey: C III Emission Lines in Of Spectra}",
      journal = {\apjl},
         year = 2010,
        month = mar,
       volume = {711},
       number = {2},
        pages = {L143-L147},
          doi = {10.1088/2041-8205/711/2/L143},
archivePrefix = {arXiv},
       eprint = {1002.3293},
 primaryClass = {astro-ph.SR},
       adsurl = {https://ui.adsabs.harvard.edu/abs/2010ApJ...711L.143W}
}

@ARTICLE{Dionne2006,
       author = {{Dionne}, Dany and {Robert}, Carmelle},
        title = "{Evolutionary Synthesis Models of Young Star-forming Regions: The Influence of Binary Stars}",
      journal = {\apj},
         year = 2006,
        month = apr,
       volume = {641},
       number = {1},
        pages = {252-267},
          doi = {10.1086/500380},
       adsurl = {https://ui.adsabs.harvard.edu/abs/2006ApJ...641..252D}
}

@ARTICLE{Hovis2025,
       author = {{Hovis-Afflerbach}, B. and {G{\"o}tberg}, Y. and {Schootemeijer}, A. and {Klencki}, J. and {Strom}, A.~L. and {Ludwig}, B.~A. and {Drout}, M.~R.},
        title = "{The mass distribution of stars stripped in binaries: The effect of metallicity}",
      journal = {\aap},
         year = 2025,
        month = may,
       volume = {697},
          eid = {A239},
        pages = {A239},
          doi = {10.1051/0004-6361/202453185},
archivePrefix = {arXiv},
       eprint = {2412.05356},
 primaryClass = {astro-ph.SR},
       adsurl = {https://ui.adsabs.harvard.edu/abs/2025A&A...697A.239H}
}

@ARTICLE{Yungelson2024,
       author = {{Yungelson}, L. and {Kuranov}, A. and {Postnov}, K. and {Kuranova}, M. and {Oskinova}, L.~M. and {Hamann}, W.-R.},
        title = "{Elusive hot stripped helium stars in the Galaxy. I. Evolutionary stellar models in the gap between subdwarfs and Wolf-Rayet stars}",
      journal = {\aap},
         year = 2024,
        month = mar,
       volume = {683},
          eid = {A37},
        pages = {A37},
          doi = {10.1051/0004-6361/202347806},
       adsurl = {https://ui.adsabs.harvard.edu/abs/2024A&A...683A..37Y}
}

@ARTICLE{Oskinova2005,
       author = {{Oskinova}, L.~M.},
        title = "{Evolution of X-ray emission from young massive star clusters}",
      journal = {\mnras},
         year = 2005,
        month = aug,
       volume = {361},
       number = {2},
        pages = {679-694},
          doi = {10.1111/j.1365-2966.2005.09229.x},
archivePrefix = {arXiv},
       eprint = {astro-ph/0505512},
 primaryClass = {astro-ph},
       adsurl = {https://ui.adsabs.harvard.edu/abs/2005MNRAS.361..679O}
}

@ARTICLE{brott+2011,
       author = {{Brott}, I. and {de Mink}, S.~E. and {Cantiello}, M. and {Langer}, N. and {de Koter}, A. and {Evans}, C.~J. and {Hunter}, I. and {Trundle}, C. and {Vink}, J.~S.},
        title = "{Rotating massive main-sequence stars. I. Grids of evolutionary models and isochrones}",
      journal = {\aap},
         year = 2011,
        month = jun,
       volume = {530},
          eid = {A115},
        pages = {A115},
          doi = {10.1051/0004-6361/201016113},
archivePrefix = {arXiv},
       eprint = {1102.0530},
 primaryClass = {astro-ph.SR},
       adsurl = {https://ui.adsabs.harvard.edu/abs/2011A&A...530A.115B}
}

@ARTICLE{pauli+2025,
       author = {{Pauli}, D. and {Oskinova}, L.~M. and {Hamann}, W.-R. and {Sander}, A.~A.~C. and {Vink}, J.~S. and {Bernini-Peron}, M. and {Josiek}, J. and {Lefever}, R.~R. and {Sana}, H. and {Ramachandran}, V.},
        title = "{New empirical mass-loss recipe for UV radiation line-driven winds of hot stars across various metallicities}",
      journal = {\aap},
         year = 2025,
        month = may,
       volume = {697},
          eid = {A114},
        pages = {A114},
          doi = {10.1051/0004-6361/202553910},
archivePrefix = {arXiv},
       eprint = {2504.07073},
 primaryClass = {astro-ph.SR},
       adsurl = {https://ui.adsabs.harvard.edu/abs/2025A&A...697A.114P}
}

@ARTICLE{Eldridge2017,
       author = {{Eldridge}, J.~J. and {Stanway}, E.~R. and {Xiao}, L. and {McClelland}, L.~A.~S. and {Taylor}, G. and {Ng}, M. and {Greis}, S.~M.~L. and {Bray}, J.~C.},
        title = "{Binary Population and Spectral Synthesis Version 2.1: Construction, Observational Verification, and New Results}",
      journal = {\pasa},
         year = 2017,
        month = nov,
       volume = {34},
          eid = {e058},
        pages = {e058},
          doi = {10.1017/pasa.2017.51},
archivePrefix = {arXiv},
       eprint = {1710.02154},
 primaryClass = {astro-ph.SR},
       adsurl = {https://ui.adsabs.harvard.edu/abs/2017PASA...34...58E}
}

@ARTICLE{Stanway&Eldridge2018,
       author = {{Stanway}, E.~R. and {Eldridge}, J.~J.},
        title = "{Re-evaluating old stellar populations}",
      journal = {\mnras},
         year = 2018,
        month = sep,
       volume = {479},
       number = {1},
        pages = {75-93},
          doi = {10.1093/mnras/sty1353},
archivePrefix = {arXiv},
       eprint = {1805.08784},
 primaryClass = {astro-ph.GA},
       adsurl = {https://ui.adsabs.harvard.edu/abs/2018MNRAS.479...75S}
}

@ARTICLE{HD45166_Picco2026,
       author = {{Picco}, A. and {Marchant}, P. and {Pauli}, D. and {Sana}, H.},
        title = "{Mergers via failed common envelope as a route toward intermediate-mass stripped stars}",
      journal = {\aap},
         year = 2026,
        month = jun,
       volume = {710},
          eid = {L12},
        pages = {L12},
          doi = {10.1051/0004-6361/202659889},
archivePrefix = {arXiv},
       eprint = {2605.22911},
 primaryClass = {astro-ph.SR},
       adsurl = {https://ui.adsabs.harvard.edu/abs/2026A&A...710L..12P}
}

@dataset{WR3_Gaia2022,
       author = {{Gaia Collaboration}},
        title = "{VizieR Online Data Catalog: Gaia DR3 Part 4. Variability (Gaia Collaboration, 2022)}",
 howpublished = {VizieR On-line Data Catalog: I/358.  Originally published in: 2023A\&A...674A...1G},
         year = 2022,
        month = may,
          eid = {I/358},
       adsurl = {https://ui.adsabs.harvard.edu/abs/2022yCat.1358....0G}
}

@ARTICLE{WRQCHE_Boco2025,
       author = {{Boco}, Lumen and {Mapelli}, Michela and {Sander}, Andreas A.~C. and {Mesini}, Sofia and {Ramachandran}, Varsha and {Torniamenti}, Stefano and {Korb}, Erika and {Liu}, Boyuan and {Sabhahit}, Gautham N. and {Vink}, Jorick S.},
        title = "{Metal-poor single Wolf-Rayet stars: The interplay of optically thick winds and rotation}",
      journal = {\aap},
         year = 2025,
        month = nov,
       volume = {703},
          eid = {A243},
        pages = {A243},
          doi = {10.1051/0004-6361/202556187},
archivePrefix = {arXiv},
       eprint = {2507.00137},
 primaryClass = {astro-ph.SR},
       adsurl = {https://ui.adsabs.harvard.edu/abs/2025A&A...703A.243B}
}

@ARTICLE{MASSLOSS_Schootemeijer2024,
       author = {{Schootemeijer}, A. and {Shenar}, T. and {Langer}, N. and {Grin}, N. and {Sana}, H. and {Gr{\"a}fener}, G. and {Sch{\"u}rmann}, C. and {Wang}, C. and {Xu}, X.-T.},
        title = "{An absence of binary companions to Wolf-Rayet stars in the Small Magellanic Cloud: Implications for mass loss and black hole masses at low metallicities}",
      journal = {\aap},
         year = 2024,
        month = sep,
       volume = {689},
          eid = {A157},
        pages = {A157},
          doi = {10.1051/0004-6361/202449978},
archivePrefix = {arXiv},
       eprint = {2406.01420},
 primaryClass = {astro-ph.SR},
       adsurl = {https://ui.adsabs.harvard.edu/abs/2024A&A...689A.157S}
}

@ARTICLE{HD45166_Shenar2023,
       author = {{Shenar}, Tomer and {Wade}, Gregg A. and {Marchant}, Pablo and {Bagnulo}, Stefano and {Bodensteiner}, Julia and {Bowman}, Dominic M. and {Gilkis}, Avishai and {Langer}, Norbert and {Nicolas-Chen{\'e}}, Andr{\'e} and {Oskinova}, Lidia and {Van Reeth}, Timothy and {Sana}, Hugues and {St-Louis}, Nicole and {de Oliveira}, Alexandre Soares and {Todt}, Helge and {Toonen}, Silvia},
        title = "{A massive helium star with a sufficiently strong magnetic field to form a magnetar}",
      journal = {Science},
         year = 2023,
        month = aug,
       volume = {381},
       number = {6659},
        pages = {761-765},
          doi = {10.1126/science.ade3293},
archivePrefix = {arXiv},
       eprint = {2308.08591},
 primaryClass = {astro-ph.SR},
       adsurl = {https://ui.adsabs.harvard.edu/abs/2023Sci...381..761S}
}

@ARTICLE{STRIPPED_DroutGotberg2023,
       author = {{Drout}, M.~R. and {G{\"o}tberg}, Y. and {Ludwig}, B.~A. and {Groh}, J.~H. and {de Mink}, S.~E. and {O'Grady}, A.~J.~G. and {Smith}, N.},
        title = "{An observed population of intermediate-mass helium stars that have been stripped in binaries}",
      journal = {Science},
         year = 2023,
        month = dec,
       volume = {382},
       number = {6676},
        pages = {1287-1291},
          doi = {10.1126/science.ade4970},
archivePrefix = {arXiv},
       eprint = {2307.00061},
 primaryClass = {astro-ph.SR},
       adsurl = {https://ui.adsabs.harvard.edu/abs/2023Sci...382.1287D}
}

@ARTICLE{OFP_Walborn1972,
       author = {{Walborn}, N.~R.},
        title = "{Spectral classification of OB stars in both hemispheres and the absolute-magnitude calibration.}",
      journal = {\aj},
         year = 1972,
        month = may,
       volume = {77},
        pages = {312-318},
          doi = {10.1086/111285},
       adsurl = {https://ui.adsabs.harvard.edu/abs/1972AJ.....77..312W}
}

@ARTICLE{WR_Shenar2020a,
       author = {{Shenar}, T. and {Gilkis}, A. and {Vink}, J.~S. and {Sana}, H. and {Sander}, A.~A.~C.},
        title = "{Why binary interaction does not necessarily dominate the formation of Wolf-Rayet stars at low metallicity}",
      journal = {\aap},
         year = 2020,
        month = feb,
       volume = {634},
          eid = {A79},
        pages = {A79},
          doi = {10.1051/0004-6361/201936948},
archivePrefix = {arXiv},
       eprint = {2001.04476},
 primaryClass = {astro-ph.SR},
       adsurl = {https://ui.adsabs.harvard.edu/abs/2020A&A...634A..79S}
}

@ARTICLE{CMFGEN_Hillier1998,
       author = {{Hillier}, D. John and {Miller}, D.~L.},
        title = "{The Treatment of Non-LTE Line Blanketing in Spherically Expanding Outflows}",
      journal = {\apj},
         year = 1998,
        month = mar,
       volume = {496},
       number = {1},
        pages = {407-427},
          doi = {10.1086/305350},
       adsurl = {https://ui.adsabs.harvard.edu/abs/1998ApJ...496..407H}
}

@ARTICLE{BINARIES_vanderHucht2001,
       author = {{van der Hucht}, Karel A.},
        title = "{The VIIth catalogue of galactic Wolf-Rayet stars}",
      journal = {\nar},
         year = 2001,
        month = feb,
       volume = {45},
       number = {3},
        pages = {135-232},
          doi = {10.1016/S1387-6473(00)00112-3},
       adsurl = {https://ui.adsabs.harvard.edu/abs/2001NewAR..45..135V}
}

@ARTICLE{SMCWR_Aguilera-Dena2022,
       author = {{Aguilera-Dena}, David R. and {Langer}, Norbert and {Antoniadis}, John and {Pauli}, Daniel and {Dessart}, Luc and {Vigna-G{\'o}mez}, Alejandro and {Gr{\"a}fener}, G{\"o}tz and {Yoon}, Sung-Chul},
        title = "{Stripped-envelope stars in different metallicity environments. I. Evolutionary phases, classification, and populations}",
      journal = {\aap},
         year = 2022,
        month = may,
       volume = {661},
          eid = {A60},
        pages = {A60},
          doi = {10.1051/0004-6361/202142895},
archivePrefix = {arXiv},
       eprint = {2112.06948},
 primaryClass = {astro-ph.SR},
       adsurl = {https://ui.adsabs.harvard.edu/abs/2022A&A...661A..60A}
}

@ARTICLE{SMCWR_Ramachandran2019,
       author = {{Ramachandran}, V. and {Hamann}, W.-R. and {Oskinova}, L.~M. and {Gallagher}, J.~S. and {Hainich}, R. and {Shenar}, T. and {Sander}, A.~A.~C. and {Todt}, H. and {Fulmer}, L.},
        title = "{Testing massive star evolution, star formation history, and feedback at low metallicity. Spectroscopic analysis of OB stars in the SMC Wing}",
      journal = {\aap},
         year = 2019,
        month = may,
       volume = {625},
          eid = {A104},
        pages = {A104},
          doi = {10.1051/0004-6361/201935365},
archivePrefix = {arXiv},
       eprint = {1903.01762},
 primaryClass = {astro-ph.SR},
       adsurl = {https://ui.adsabs.harvard.edu/abs/2019A&A...625A.104R}
}

@ARTICLE{SMCWR_Schootemeijer2018,
       author = {{Schootemeijer}, A. and {Langer}, N.},
        title = "{Wolf-Rayet stars in the Small Magellanic Cloud as testbed for massive star evolution}",
      journal = {\aap},
         year = 2018,
        month = mar,
       volume = {611},
          eid = {A75},
        pages = {A75},
          doi = {10.1051/0004-6361/201731895},
archivePrefix = {arXiv},
       eprint = {1709.08727},
 primaryClass = {astro-ph.SR},
       adsurl = {https://ui.adsabs.harvard.edu/abs/2018A&A...611A..75S}
}

@ARTICLE{SMCWR_Martins2009,
       author = {{Martins}, F. and {Hillier}, D.~J. and {Bouret}, J.~C. and {Depagne}, E. and {Foellmi}, C. and {Marchenko}, S. and {Moffat}, A.~F.},
        title = "{Properties of WNh stars in the Small Magellanic Cloud: evidence for homogeneous evolution}",
      journal = {\aap},
         year = 2009,
        month = feb,
       volume = {495},
       number = {1},
        pages = {257-270},
          doi = {10.1051/0004-6361:200811014},
archivePrefix = {arXiv},
       eprint = {0811.3564},
 primaryClass = {astro-ph},
       adsurl = {https://ui.adsabs.harvard.edu/abs/2009A&A...495..257M}
}

@ARTICLE{SMCSURVEY_Massey2014,
       author = {{Massey}, Philip and {Neugent}, Kathryn F. and {Morrell}, Nidia and {Hillier}, D. John},
        title = "{A Modern Search for Wolf-Rayet Stars in the Magellanic Clouds: First Results}",
      journal = {\apj},
         year = 2014,
        month = jun,
       volume = {788},
       number = {1},
          eid = {83},
        pages = {83},
          doi = {10.1088/0004-637X/788/1/83},
archivePrefix = {arXiv},
       eprint = {1404.7441},
 primaryClass = {astro-ph.SR},
       adsurl = {https://ui.adsabs.harvard.edu/abs/2014ApJ...788...83M}
}

@ARTICLE{SMCSURVEY_Neugent2018,
       author = {{Neugent}, Kathryn F. and {Massey}, Philip and {Morrell}, Nidia},
        title = "{A Modern Search for Wolf-Rayet Stars in the Magellanic Clouds. IV. A Final Census}",
      journal = {\apj},
         year = 2018,
        month = aug,
       volume = {863},
       number = {2},
          eid = {181},
        pages = {181},
          doi = {10.3847/1538-4357/aad17d},
archivePrefix = {arXiv},
       eprint = {1807.01209},
 primaryClass = {astro-ph.SR},
       adsurl = {https://ui.adsabs.harvard.edu/abs/2018ApJ...863..181N}
}

@ARTICLE{MINLUM_Sander2020,
       author = {{Sander}, Andreas A.~C. and {Vink}, Jorick S.},
        title = "{On the nature of massive helium star winds and Wolf-Rayet-type mass-loss}",
      journal = {\mnras},
         year = 2020,
        month = nov,
       volume = {499},
       number = {1},
        pages = {873-892},
          doi = {10.1093/mnras/staa2712},
archivePrefix = {arXiv},
       eprint = {2009.01849},
 primaryClass = {astro-ph.SR},
       adsurl = {https://ui.adsabs.harvard.edu/abs/2020MNRAS.499..873S}
}

@ARTICLE{BINARIES_Eldridge2022,
       author = {{Eldridge}, Jan J. and {Stanway}, Elizabeth R.},
        title = "{New Insights into the Evolution of Massive Stars and Their Effects on Our Understanding of Early Galaxies}",
      journal = {\araa},
         year = 2022,
        month = aug,
       volume = {60},
        pages = {455-494},
          doi = {10.1146/annurev-astro-052920-100646},
archivePrefix = {arXiv},
       eprint = {2202.01413},
 primaryClass = {astro-ph.GA},
       adsurl = {https://ui.adsabs.harvard.edu/abs/2022ARA&A..60..455E}
}

@ARTICLE{SMCMET_Pilyugin2001,
       author = {{Pilyugin}, L.~S.},
        title = "{Oxygen abundances in dwarf irregular galaxies and the metallicity-luminosity relationship}",
      journal = {\aap},
         year = 2001,
        month = aug,
       volume = {374},
        pages = {412-420},
          doi = {10.1051/0004-6361:20010732},
archivePrefix = {arXiv},
       eprint = {astro-ph/0105360},
 primaryClass = {astro-ph},
       adsurl = {https://ui.adsabs.harvard.edu/abs/2001A&A...374..412P}
}

@ARTICLE{POWR_Todt2015,
       author = {{Todt}, H. and {Sander}, A. and {Hainich}, R. and {Hamann}, W.-R. and {Quade}, M. and {Shenar}, T.},
        title = "{Potsdam Wolf-Rayet model atmosphere grids for WN stars}",
      journal = {\aap},
         year = 2015,
        month = jul,
       volume = {579},
          eid = {A75},
        pages = {A75},
          doi = {10.1051/0004-6361/201526253},
       adsurl = {https://ui.adsabs.harvard.edu/abs/2015A&A...579A..75T}
}

@ARTICLE{REDDENING_Schlafly2011,
       author = {{Schlafly}, Edward F. and {Finkbeiner}, Douglas P.},
        title = "{Measuring Reddening with Sloan Digital Sky Survey Stellar Spectra and Recalibrating SFD}",
      journal = {\apj},
         year = 2011,
        month = aug,
       volume = {737},
       number = {2},
          eid = {103},
        pages = {103},
          doi = {10.1088/0004-637X/737/2/103},
archivePrefix = {arXiv},
       eprint = {1012.4804},
 primaryClass = {astro-ph.GA},
       adsurl = {https://ui.adsabs.harvard.edu/abs/2011ApJ...737..103S}
}

@ARTICLE{MAGNETICWR_Hubrig2016,
       author = {{Hubrig}, S. and {Scholz}, K. and {Hamann}, W.-R. and {Sch{\"o}ller}, M. and {Ignace}, R. and {Ilyin}, I. and {Gayley}, K.~G. and {Oskinova}, L.~M.},
        title = "{Searching for a magnetic field in Wolf-Rayet stars using FORS 2 spectropolarimetry}",
      journal = {\mnras},
         year = 2016,
        month = may,
       volume = {458},
       number = {3},
        pages = {3381-3393},
          doi = {10.1093/mnras/stw558},
archivePrefix = {arXiv},
       eprint = {1603.01441},
 primaryClass = {astro-ph.SR},
       adsurl = {https://ui.adsabs.harvard.edu/abs/2016MNRAS.458.3381H}
}

@ARTICLE{CLUMPING_Hamann1998,
       author = {{Hamann}, W.-R. and {Koesterke}, L.},
        title = "{Spectrum formation in clumped stellar winds: consequences for the analyses of Wolf-Rayet spectra}",
      journal = {\aap},
         year = 1998,
        month = jul,
       volume = {335},
        pages = {1003-1008},
       adsurl = {https://ui.adsabs.harvard.edu/abs/1998A&A...335.1003H}
}

@PHDTHESIS{WR12_Abbot2004,
       author = {{Abbott}, Jay Brian},
        title = "{Quantitative spectroscopic studies of Wolf-Rayet stars in local group galaxies}",
       school = {University College London, UK},
         year = 2004,
        month = jan,
       adsurl = {https://ui.adsabs.harvard.edu/abs/2004PhDT.......161A}
}

@ARTICLE{MASSLUM_Grafener2011,
       author = {{Gr{\"a}fener}, G. and {Vink}, J.~S. and {de Koter}, A. and {Langer}, N.},
        title = "{The Eddington factor as the key to understand the winds of the most massive stars. Evidence for a {\ensuremath{\Gamma}}-dependence of Wolf-Rayet type mass loss}",
      journal = {\aap},
         year = 2011,
        month = nov,
       volume = {535},
          eid = {A56},
        pages = {A56},
          doi = {10.1051/0004-6361/201116701},
archivePrefix = {arXiv},
       eprint = {1106.5361},
 primaryClass = {astro-ph.SR},
       adsurl = {https://ui.adsabs.harvard.edu/abs/2011A&A...535A..56G}
}

@ARTICLE{HCOLMW_Groenewegen1989,
       author = {{Groenewegen}, M.~A.~T. and {Lamers}, H.~J.~G.~L.~M.},
        title = "{The winds of O-stars. I. an analysis of the UV line profiles with theSEI method.}",
      journal = {\aaps},
         year = 1989,
        month = sep,
       volume = {79},
        pages = {359-383},
       adsurl = {https://ui.adsabs.harvard.edu/abs/1989A&AS...79..359G}
}

@ARTICLE{CLUMPING_Najarro2009,
       author = {{Najarro}, Francisco and {Figer}, Don F. and {Hillier}, D. John and {Geballe}, T.~R. and {Kudritzki}, Rolf P.},
        title = "{Metallicity in the Galactic Center: The Quintuplet Cluster}",
      journal = {\apj},
         year = 2009,
        month = feb,
       volume = {691},
       number = {2},
        pages = {1816-1827},
          doi = {10.1088/0004-637X/691/2/1816},
archivePrefix = {arXiv},
       eprint = {0809.3185},
 primaryClass = {astro-ph},
       adsurl = {https://ui.adsabs.harvard.edu/abs/2009ApJ...691.1816N}
}

@ARTICLE{WR12_Bessel1998,
       author = {{Bessell}, M.~S. and {Castelli}, F. and {Plez}, B.},
        title = "{Model atmospheres broad-band colors, bolometric corrections and temperature calibrations for O - M stars}",
      journal = {\aap},
         year = 1998,
        month = may,
       volume = {333},
        pages = {231-250},
       adsurl = {https://ui.adsabs.harvard.edu/abs/1998A&A...333..231B}
}

@ARTICLE{REDDENING_Cardelli1989,
       author = {{Cardelli}, Jason A. and {Clayton}, Geoffrey C. and {Mathis}, John S.},
        title = "{The Relationship between Infrared, Optical, and Ultraviolet Extinction}",
      journal = {\apj},
         year = 1989,
        month = oct,
       volume = {345},
        pages = {245},
          doi = {10.1086/167900},
       adsurl = {https://ui.adsabs.harvard.edu/abs/1989ApJ...345..245C}
}

@ARTICLE{REDDENING_Seaton1979,
       author = {{Seaton}, M.~J.},
        title = "{Interstellar extinction in the UV.}",
      journal = {\mnras},
         year = 1979,
        month = jan,
       volume = {187},
        pages = {73},
          doi = {10.1093/mnras/187.1.73P},
       adsurl = {https://ui.adsabs.harvard.edu/abs/1979MNRAS.187P..73S}
}

@ARTICLE{BETA_Grafener2005,
       author = {{Gr{\"a}fener}, G. and {Hamann}, W.-R.},
        title = "{Hydrodynamic model atmospheres for WR stars. Self-consistent modeling of a WC star wind}",
      journal = {\aap},
         year = 2005,
        month = mar,
       volume = {432},
       number = {2},
        pages = {633-645},
          doi = {10.1051/0004-6361:20041732},
archivePrefix = {arXiv},
       eprint = {astro-ph/0410697},
 primaryClass = {astro-ph},
       adsurl = {https://ui.adsabs.harvard.edu/abs/2005A&A...432..633G}
}

@ARTICLE{BETA_Hillier1999,
       author = {{Hillier}, D. John and {Miller}, D.~L.},
        title = "{Constraints on the Evolution of Massive Stars through Spectral Analysis. I. The WC5 Star HD 165763}",
      journal = {\apj},
         year = 1999,
        month = jul,
       volume = {519},
       number = {1},
        pages = {354-371},
          doi = {10.1086/307339},
       adsurl = {https://ui.adsabs.harvard.edu/abs/1999ApJ...519..354H}
}

@ARTICLE{BETA_Pauldrach1986,
       author = {{Pauldrach}, A. and {Puls}, J. and {Kudritzki}, R.~P.},
        title = "{Radiation-driven winds of hot luminous stars. Improvements of the theory and first results.}",
      journal = {\aap},
         year = 1986,
        month = aug,
       volume = {164},
        pages = {86-100},
       adsurl = {https://ui.adsabs.harvard.edu/abs/1986A&A...164...86P}
}

@ARTICLE{BETA_Castor1979,
       author = {{Castor}, J.~I. and {Lamers}, H.~J.~G.~L.~M.},
        title = "{An atlas of theoretical P Cygni profiles.}",
      journal = {\apjs},
         year = 1979,
        month = apr,
       volume = {39},
        pages = {481-511},
          doi = {10.1086/190583},
       adsurl = {https://ui.adsabs.harvard.edu/abs/1979ApJS...39..481C}
}

@ARTICLE{POWR_Sander2015,
       author = {{Sander}, A. and {Shenar}, T. and {Hainich}, R. and {G{\'\i}menez-Garc{\'\i}a}, A. and {Todt}, H. and {Hamann}, W.-R.},
        title = "{On the consistent treatment of the quasi-hydrostatic layers in hot star atmospheres}",
      journal = {\aap},
         year = 2015,
        month = may,
       volume = {577},
          eid = {A13},
        pages = {A13},
          doi = {10.1051/0004-6361/201425356},
archivePrefix = {arXiv},
       eprint = {1503.01338},
 primaryClass = {astro-ph.SR},
       adsurl = {https://ui.adsabs.harvard.edu/abs/2015A&A...577A..13S}
}

@ARTICLE{POWR_Hamann2004,
       author = {{Hamann}, W.-R. and {Gr{\"a}fener}, G.},
        title = "{Grids of model spectra for WN stars, ready for use}",
      journal = {\aap},
         year = 2004,
        month = nov,
       volume = {427},
        pages = {697-704},
          doi = {10.1051/0004-6361:20040506},
       adsurl = {https://ui.adsabs.harvard.edu/abs/2004A&A...427..697H}
}

@ARTICLE{POWR_Hamann2003,
       author = {{Hamann}, W.-R. and {Gr{\"a}fener}, G.},
        title = "{A temperature correction method for expanding atmospheres}",
      journal = {\aap},
         year = 2003,
        month = nov,
       volume = {410},
        pages = {993-1000},
          doi = {10.1051/0004-6361:20031308},
       adsurl = {https://ui.adsabs.harvard.edu/abs/2003A&A...410..993H}
}

@ARTICLE{GAIADR3_2016,
       author = {{Gaia Collaboration} and {Prusti}, T. and {de Bruijne}, J.~H.~J. and {Brown}, A.~G.~A. and {Vallenari}, A. and {Babusiaux}, C. and {Bailer-Jones}, C.~A.~L. and {Bastian}, U. and {Biermann}, M. and {Evans}, D.~W. and {Eyer}, L. and {Jansen}, F. and {Jordi}, C. and {Klioner}, S.~A. and {Lammers}, U. and {Lindegren}, L. and {Luri}, X. and {Mignard}, F. and {Milligan}, D.~J. and {Panem}, C. and {Poinsignon}, V. and {Pourbaix}, D. and {Randich}, S. and {Sarri}, G. and {Sartoretti}, P. and {Siddiqui}, H.~I. and {Soubiran}, C. and {Valette}, V. and {van Leeuwen}, F. and {Walton}, N.~A. and {Aerts}, C. and {Arenou}, F. and {Cropper}, M. and {Drimmel}, R. and {H{\o}g}, E. and {Katz}, D. and {Lattanzi}, M.~G. and {O'Mullane}, W. and {Grebel}, E.~K. and {Holland}, A.~D. and {Huc}, C. and {Passot}, X. and {Bramante}, L. and {Cacciari}, C. and {Casta{\~n}eda}, J. and {Chaoul}, L. and {Cheek}, N. and {De Angeli}, F. and {Fabricius}, C. and {Guerra}, R. and {Hern{\'a}ndez}, J. and {Jean-Antoine-Piccolo}, A. and {Masana}, E. and {Messineo}, R. and {Mowlavi}, N. and {Nienartowicz}, K. and {Ord{\'o}{\~n}ez-Blanco}, D. and {Panuzzo}, P. and {Portell}, J. and {Richards}, P.~J. and {Riello}, M. and {Seabroke}, G.~M. and {Tanga}, P. and {Th{\'e}venin}, F. and {Torra}, J. and {Els}, S.~G. and {Gracia-Abril}, G. and {Comoretto}, G. and {Garcia-Reinaldos}, M. and {Lock}, T. and {Mercier}, E. and {Altmann}, M. and {Andrae}, R. and {Astraatmadja}, T.~L. and {Bellas-Velidis}, I. and {Benson}, K. and {Berthier}, J. and {Blomme}, R. and {Busso}, G. and {Carry}, B. and {Cellino}, A. and {Clementini}, G. and {Cowell}, S. and {Creevey}, O. and {Cuypers}, J. and {Davidson}, M. and {De Ridder}, J. and {de Torres}, A. and {Delchambre}, L. and {Dell'Oro}, A. and {Ducourant}, C. and {Fr{\'e}mat}, Y. and {Garc{\'\i}a-Torres}, M. and {Gosset}, E. and {Halbwachs}, J.-L. and {Hambly}, N.~C. and {Harrison}, D.~L. and {Hauser}, M. and {Hestroffer}, D. and {Hodgkin}, S.~T. and {Huckle}, H.~E. and {Hutton}, A. and {Jasniewicz}, G. and {Jordan}, S. and {Kontizas}, M. and {Korn}, A.~J. and {Lanzafame}, A.~C. and {Manteiga}, M. and {Moitinho}, A. and {Muinonen}, K. and {Osinde}, J. and {Pancino}, E. and {Pauwels}, T. and {Petit}, J.-M. and {Recio-Blanco}, A. and {Robin}, A.~C. and {Sarro}, L.~M. and {Siopis}, C. and {Smith}, M. and {Smith}, K.~W. and {Sozzetti}, A. and {Thuillot}, W. and {van Reeven}, W. and {Viala}, Y. and {Abbas}, U. and {Abreu Aramburu}, A. and {Accart}, S. and {Aguado}, J.~J. and {Allan}, P.~M. and {Allasia}, W. and {Altavilla}, G. and {{\'A}lvarez}, M.~A. and {Alves}, J. and {Anderson}, R.~I. and {Andrei}, A.~H. and {Anglada Varela}, E. and {Antiche}, E. and {Antoja}, T. and {Ant{\'o}n}, S. and {Arcay}, B. and {Atzei}, A. and {Ayache}, L. and {Bach}, N. and {Baker}, S.~G. and {Balaguer-N{\'u}{\~n}ez}, L. and {Barache}, C. and {Barata}, C. and {Barbier}, A. and {Barblan}, F. and {Baroni}, M. and {Barrado y Navascu{\'e}s}, D. and {Barros}, M. and {Barstow}, M.~A. and {Becciani}, U. and {Bellazzini}, M. and {Bellei}, G. and {Bello Garc{\'\i}a}, A. and {Belokurov}, V. and {Bendjoya}, P. and {Berihuete}, A. and {Bianchi}, L. and {Bienaym{\'e}}, O. and {Billebaud}, F. and {Blagorodnova}, N. and {Blanco-Cuaresma}, S. and {Boch}, T. and {Bombrun}, A. and {Borrachero}, R. and {Bouquillon}, S. and {Bourda}, G. and {Bouy}, H. and {Bragaglia}, A. and {Breddels}, M.~A. and {Brouillet}, N. and {Br{\"u}semeister}, T. and {Bucciarelli}, B. and {Budnik}, F. and {Burgess}, P. and {Burgon}, R. and {Burlacu}, A. and {Busonero}, D. and {Buzzi}, R. and {Caffau}, E. and {Cambras}, J. and {Campbell}, H. and {Cancelliere}, R. and {Cantat-Gaudin}, T. and {Carlucci}, T. and {Carrasco}, J.~M. and {Castellani}, M. and {Charlot}, P. and {Charnas}, J. and {Charvet}, P. and {Chassat}, F. and {Chiavassa}, A. and {Clotet}, M. and {Cocozza}, G. and {Collins}, R.~S. and {Collins}, P. and {Costigan}, G.},
        title = "{The Gaia mission}",
      journal = {\aap},
         year = 2016,
        month = nov,
       volume = {595},
          eid = {A1},
        pages = {A1},
          doi = {10.1051/0004-6361/201629272},
archivePrefix = {arXiv},
       eprint = {1609.04153},
 primaryClass = {astro-ph.IM},
       adsurl = {https://ui.adsabs.harvard.edu/abs/2016A&A...595A...1G}
}

@ARTICLE{GAIADR3_2023,
       author = {{Gaia Collaboration} and {Vallenari}, A. and {Brown}, A.~G.~A. and {Prusti}, T. and {de Bruijne}, J.~H.~J. and {Arenou}, F. and {Babusiaux}, C. and {Biermann}, M. and {Creevey}, O.~L. and {Ducourant}, C. and {Evans}, D.~W. and {Eyer}, L. and {Guerra}, R. and {Hutton}, A. and {Jordi}, C. and {Klioner}, S.~A. and {Lammers}, U.~L. and {Lindegren}, L. and {Luri}, X. and {Mignard}, F. and {Panem}, C. and {Pourbaix}, D. and {Randich}, S. and {Sartoretti}, P. and {Soubiran}, C. and {Tanga}, P. and {Walton}, N.~A. and {Bailer-Jones}, C.~A.~L. and {Bastian}, U. and {Drimmel}, R. and {Jansen}, F. and {Katz}, D. and {Lattanzi}, M.~G. and {van Leeuwen}, F. and {Bakker}, J. and {Cacciari}, C. and {Casta{\~n}eda}, J. and {De Angeli}, F. and {Fabricius}, C. and {Fouesneau}, M. and {Fr{\'e}mat}, Y. and {Galluccio}, L. and {Guerrier}, A. and {Heiter}, U. and {Masana}, E. and {Messineo}, R. and {Mowlavi}, N. and {Nicolas}, C. and {Nienartowicz}, K. and {Pailler}, F. and {Panuzzo}, P. and {Riclet}, F. and {Roux}, W. and {Seabroke}, G.~M. and {Sordo}, R. and {Th{\'e}venin}, F. and {Gracia-Abril}, G. and {Portell}, J. and {Teyssier}, D. and {Altmann}, M. and {Andrae}, R. and {Audard}, M. and {Bellas-Velidis}, I. and {Benson}, K. and {Berthier}, J. and {Blomme}, R. and {Burgess}, P.~W. and {Busonero}, D. and {Busso}, G. and {C{\'a}novas}, H. and {Carry}, B. and {Cellino}, A. and {Cheek}, N. and {Clementini}, G. and {Damerdji}, Y. and {Davidson}, M. and {de Teodoro}, P. and {Nu{\~n}ez Campos}, M. and {Delchambre}, L. and {Dell'Oro}, A. and {Esquej}, P. and {Fern{\'a}ndez-Hern{\'a}ndez}, J. and {Fraile}, E. and {Garabato}, D. and {Garc{\'\i}a-Lario}, P. and {Gosset}, E. and {Haigron}, R. and {Halbwachs}, J.-L. and {Hambly}, N.~C. and {Harrison}, D.~L. and {Hern{\'a}ndez}, J. and {Hestroffer}, D. and {Hodgkin}, S.~T. and {Holl}, B. and {Jan{\ss}en}, K. and {Jevardat de Fombelle}, G. and {Jordan}, S. and {Krone-Martins}, A. and {Lanzafame}, A.~C. and {L{\"o}ffler}, W. and {Marchal}, O. and {Marrese}, P.~M. and {Moitinho}, A. and {Muinonen}, K. and {Osborne}, P. and {Pancino}, E. and {Pauwels}, T. and {Recio-Blanco}, A. and {Reyl{\'e}}, C. and {Riello}, M. and {Rimoldini}, L. and {Roegiers}, T. and {Rybizki}, J. and {Sarro}, L.~M. and {Siopis}, C. and {Smith}, M. and {Sozzetti}, A. and {Utrilla}, E. and {van Leeuwen}, M. and {Abbas}, U. and {{\'A}brah{\'a}m}, P. and {Abreu Aramburu}, A. and {Aerts}, C. and {Aguado}, J.~J. and {Ajaj}, M. and {Aldea-Montero}, F. and {Altavilla}, G. and {{\'A}lvarez}, M.~A. and {Alves}, J. and {Anders}, F. and {Anderson}, R.~I. and {Anglada Varela}, E. and {Antoja}, T. and {Baines}, D. and {Baker}, S.~G. and {Balaguer-N{\'u}{\~n}ez}, L. and {Balbinot}, E. and {Balog}, Z. and {Barache}, C. and {Barbato}, D. and {Barros}, M. and {Barstow}, M.~A. and {Bartolom{\'e}}, S. and {Bassilana}, J.-L. and {Bauchet}, N. and {Becciani}, U. and {Bellazzini}, M. and {Berihuete}, A. and {Bernet}, M. and {Bertone}, S. and {Bianchi}, L. and {Binnenfeld}, A. and {Blanco-Cuaresma}, S. and {Blazere}, A. and {Boch}, T. and {Bombrun}, A. and {Bossini}, D. and {Bouquillon}, S. and {Bragaglia}, A. and {Bramante}, L. and {Breedt}, E. and {Bressan}, A. and {Brouillet}, N. and {Brugaletta}, E. and {Bucciarelli}, B. and {Burlacu}, A. and {Butkevich}, A.~G. and {Buzzi}, R. and {Caffau}, E. and {Cancelliere}, R. and {Cantat-Gaudin}, T. and {Carballo}, R. and {Carlucci}, T. and {Carnerero}, M.~I. and {Carrasco}, J.~M. and {Casamiquela}, L. and {Castellani}, M. and {Castro-Ginard}, A. and {Chaoul}, L. and {Charlot}, P. and {Chemin}, L. and {Chiaramida}, V. and {Chiavassa}, A. and {Chornay}, N. and {Comoretto}, G. and {Contursi}, G. and {Cooper}, W.~J. and {Cornez}, T. and {Cowell}, S. and {Crifo}, F. and {Cropper}, M. and {Crosta}, M. and {Crowley}, C. and {Dafonte}, C. and {Dapergolas}, A. and {David}, M. and {David}, P. and {de Laverny}, P. and {De Luise}, F. and {De March}, R.},
        title = "{Gaia Data Release 3. Summary of the content and survey properties}",
      journal = {\aap},
         year = 2023,
        month = jun,
       volume = {674},
          eid = {A1},
        pages = {A1},
          doi = {10.1051/0004-6361/202243940},
archivePrefix = {arXiv},
       eprint = {2208.00211},
 primaryClass = {astro-ph.GA},
       adsurl = {https://ui.adsabs.harvard.edu/abs/2023A&A...674A...1G}
}

@ARTICLE{COS_Green2012,
       author = {{Green}, James C. and {Froning}, Cynthia S. and {Osterman}, Steve and {Ebbets}, Dennis and {Heap}, Sara H. and {Leitherer}, Claus and {Linsky}, Jeffrey L. and {Savage}, Blair D. and {Sembach}, Kenneth and {Shull}, J. Michael and {Siegmund}, Oswald H.~W. and {Snow}, Theodore P. and {Spencer}, John and {Stern}, S. Alan and {Stocke}, John and {Welsh}, Barry and {B{\'e}land}, St{\'e}phane and {Burgh}, Eric B. and {Danforth}, Charles and {France}, Kevin and {Keeney}, Brian and {McPhate}, Jason and {Penton}, Steven V. and {Andrews}, John and {Brownsberger}, Kenneth and {Morse}, Jon and {Wilkinson}, Erik},
        title = "{The Cosmic Origins Spectrograph}",
      journal = {\apj},
         year = 2012,
        month = jan,
       volume = {744},
       number = {1},
          eid = {60},
        pages = {60},
          doi = {10.1088/0004-637X/744/1/6010.1086/141956},
archivePrefix = {arXiv},
       eprint = {1110.0462},
 primaryClass = {astro-ph.IM},
       adsurl = {https://ui.adsabs.harvard.edu/abs/2012ApJ...744...60G}
}

@ARTICLE{FORS2_Appenzeller1998,
       author = {{Appenzeller}, I. and {Fricke}, K. and {F{\"u}rtig}, W. and {G{\"a}ssler}, W. and {H{\"a}fner}, R. and {Harke}, R. and {Hess}, H.-J. and {Hummel}, W. and {J{\"u}rgens}, P. and {Kudritzki}, R.-P. and {Mantel}, K.-H. and {Meisl}, W. and {Muschielok}, B. and {Nicklas}, H. and {Rupprecht}, G. and {Seifert}, W. and {Stahl}, O. and {Szeifert}, T. and {Tarantik}, K.},
        title = "{Successful commissioning of FORS1 - the first optical instrument on the VLT.}",
      journal = {The Messenger},
         year = 1998,
        month = dec,
       volume = {94},
        pages = {1-6},
       adsurl = {https://ui.adsabs.harvard.edu/abs/1998Msngr..94....1A}
}

@ARTICLE{NGCMET_Hernandez-Martinez2009,
       author = {{Hern{\'a}ndez-Mart{\'\i}nez}, L. and {Pe{\~n}a}, M. and {Carigi}, L. and {Garc{\'\i}a-Rojas}, J.},
        title = "{Chemical behavior of the dwarf irregular galaxy NGC6822. Its PN and HII region abundances}",
      journal = {\aap},
         year = 2009,
        month = oct,
       volume = {505},
       number = {3},
        pages = {1027-1039},
          doi = {10.1051/0004-6361/200912476},
archivePrefix = {arXiv},
       eprint = {0906.4402},
 primaryClass = {astro-ph.CO},
       adsurl = {https://ui.adsabs.harvard.edu/abs/2009A&A...505.1027H}
}

@ARTICLE{SFH_Fusco2014,
       author = {{Fusco}, F. and {Buonanno}, R. and {Hidalgo}, S.~L. and {Aparicio}, A. and {Pietrinferni}, A. and {Bono}, G. and {Monelli}, M. and {Cassisi}, S.},
        title = "{A state-of-the-art analysis of the dwarf irregular galaxy NGC 6822}",
      journal = {\aap},
         year = 2014,
        month = dec,
       volume = {572},
          eid = {A26},
        pages = {A26},
          doi = {10.1051/0004-6361/201323075},
archivePrefix = {arXiv},
       eprint = {1409.5247},
 primaryClass = {astro-ph.GA},
       adsurl = {https://ui.adsabs.harvard.edu/abs/2014A&A...572A..26F}
}

@ARTICLE{SMCBIN_Foellmi2003,
       author = {{Foellmi}, C. and {Moffat}, A.~F.~J. and {Guerrero}, M.~A.},
        title = "{Wolf-Rayet binaries in the Magellanic Clouds and implications for massive-star evolution - I. Small Magellanic Cloud}",
      journal = {\mnras},
         year = 2003,
        month = jan,
       volume = {338},
       number = {2},
        pages = {360-388},
          doi = {10.1046/j.1365-8711.2003.06052.x},
       adsurl = {https://ui.adsabs.harvard.edu/abs/2003MNRAS.338..360F}
}

@ARTICLE{delaChevrotiere2014,
       author = {{de la Chevroti{\`e}re}, A. and {St-Louis}, N. and {Moffat}, A.~F.~J. and {MiMeS Collaboration}},
        title = "{Searching for Magnetic Fields in 11 Wolf-Rayet Stars: Analysis of Circular Polarization Measurements from ESPaDOnS}",
      journal = {\apj},
         year = 2014,
        month = feb,
       volume = {781},
       number = {2},
          eid = {73},
        pages = {73},
          doi = {10.1088/0004-637X/781/2/73},
       adsurl = {https://ui.adsabs.harvard.edu/abs/2014ApJ...781...73D}
}

@ARTICLE{Deshmukh2024,
       author = {{Deshmukh}, K. and {Sana}, H. and {M{\'e}rand}, A. and {Bordier}, E. and {Langer}, N. and {Bodensteiner}, J. and {Dsilva}, K. and {Frost}, A.~J. and {Gosset}, E. and {Le Bouquin}, J.-B. and {Lefever}, R.~R. and {Mahy}, L. and {Patrick}, L.~R. and {Reggiani}, M. and {Sander}, A.~A.~C. and {Shenar}, T. and {Tramper}, F. and {Villase{\~n}or}, J.~I. and {Waisberg}, I.},
        title = "{Investigating 39 Galactic Wolf-Rayet stars with VLTI/GRAVITY: Uncovering a long-period binary desert}",
      journal = {\aap},
         year = 2024,
        month = dec,
       volume = {692},
          eid = {A109},
        pages = {A109},
          doi = {10.1051/0004-6361/202452352},
archivePrefix = {arXiv},
       eprint = {2409.15212},
 primaryClass = {astro-ph.SR},
       adsurl = {https://ui.adsabs.harvard.edu/abs/2024A&A...692A.109D}
}

@ARTICLE{Dsilva2023,
       author = {{Dsilva}, K. and {Shenar}, T. and {Sana}, H. and {Marchant}, P.},
        title = "{A spectroscopic multiplicity survey of Galactic Wolf-Rayet stars . III. The northern late-type nitrogen-rich sample}",
      journal = {\aap},
         year = 2023,
        month = jun,
       volume = {674},
          eid = {A88},
        pages = {A88},
          doi = {10.1051/0004-6361/202244308},
archivePrefix = {arXiv},
       eprint = {2212.06927},
 primaryClass = {astro-ph.SR},
       adsurl = {https://ui.adsabs.harvard.edu/abs/2023A&A...674A..88D}
}

@ARTICLE{NGCWR_Bianchi2001a,
       author = {{Bianchi}, Luciana and {Catanzaro}, Giovanni and {Scuderi}, Salvatore and {Hutchings}, John B.},
        title = "{Spectroscopy of Massive Stars in NGC 6822 and M33}",
      journal = {\pasp},
         year = 2001,
        month = jun,
       volume = {113},
       number = {784},
        pages = {697-702},
          doi = {10.1086/320801},
       adsurl = {https://ui.adsabs.harvard.edu/abs/2001PASP..113..697B}
}

@ARTICLE{Pauli2022,
       author = {{Pauli}, D. and {Langer}, N. and {Aguilera-Dena}, D.~R. and {Wang}, C. and {Marchant}, P.},
        title = "{A synthetic population of Wolf-Rayet stars in the LMC based on detailed single and binary star evolution models}",
      journal = {\aap},
         year = 2022,
        month = nov,
       volume = {667},
          eid = {A58},
        pages = {A58},
          doi = {10.1051/0004-6361/202243965},
archivePrefix = {arXiv},
       eprint = {2208.10194},
 primaryClass = {astro-ph.SR},
       adsurl = {https://ui.adsabs.harvard.edu/abs/2022A&A...667A..58P}
}

@ARTICLE{M31M33_NeugentMassey2014,
       author = {{Neugent}, Kathryn F. and {Massey}, Philip},
        title = "{The Close Binary Frequency of Wolf-Rayet Stars as a Function of Metallicity in M31 and M33}",
      journal = {\apj},
         year = 2014,
        month = jul,
       volume = {789},
       number = {1},
          eid = {10},
        pages = {10},
          doi = {10.1088/0004-637X/789/1/10},
archivePrefix = {arXiv},
       eprint = {1405.1810},
 primaryClass = {astro-ph.SR},
       adsurl = {https://ui.adsabs.harvard.edu/abs/2014ApJ...789...10N}
}

@ARTICLE{WRBASICS_Crowther2007,
       author = {{Crowther}, Paul A.},
        title = "{Physical Properties of Wolf-Rayet Stars}",
      journal = {\araa},
         year = 2007,
        month = sep,
       volume = {45},
       number = {1},
        pages = {177-219},
          doi = {10.1146/annurev.astro.45.051806.110615},
archivePrefix = {arXiv},
       eprint = {astro-ph/0610356},
 primaryClass = {astro-ph},
       adsurl = {https://ui.adsabs.harvard.edu/abs/2007ARA&A..45..177C}
}

@ARTICLE{STRIPPED_Ramachandran2024,
       author = {{Ramachandran}, V. and {Sander}, A.~A.~C. and {Pauli}, D. and {Klencki}, J. and {Backs}, F. and {Tramper}, F. and {Bernini-Peron}, M. and {Crowther}, P. and {Hamann}, W.-R. and {Ignace}, R. and {Kuiper}, R. and {Oey}, M.~S. and {Oskinova}, L.~M. and {Shenar}, T. and {Todt}, H. and {Vink}, J.~S. and {Wang}, L. and {Wofford}, A. and {the XShootU Collaboration}},
        title = "{X-Shooting ULLYSES: Massive stars at low metallicity: VIII. Stellar and wind parameters of newly revealed stripped stars in Be binaries}",
      journal = {\aap},
         year = 2024,
        month = dec,
       volume = {692},
          eid = {A90},
        pages = {A90},
          doi = {10.1051/0004-6361/202449665},
archivePrefix = {arXiv},
       eprint = {2406.17678},
 primaryClass = {astro-ph.SR},
       adsurl = {https://ui.adsabs.harvard.edu/abs/2024A&A...692A..90R}
}

@ARTICLE{VELO_Lefever2023,
       author = {{Lefever}, R.~R. and {Sander}, A.~A.~C. and {Shenar}, T. and {Poniatowski}, L.~G. and {Dsilva}, K. and {Todt}, H.},
        title = "{Exploring the influence of different velocity fields on Wolf-Rayet star spectra}",
      journal = {\mnras},
         year = 2023,
        month = may,
       volume = {521},
       number = {1},
        pages = {1374-1392},
          doi = {10.1093/mnras/stad625},
archivePrefix = {arXiv},
       eprint = {2302.13964},
 primaryClass = {astro-ph.SR},
       adsurl = {https://ui.adsabs.harvard.edu/abs/2023MNRAS.521.1374L}
}

@ARTICLE{EVO_Martins2023,
       author = {{Martins}, Fabrice},
        title = "{Surface chemical composition of single WNh stars}",
      journal = {\aap},
         year = 2023,
        month = dec,
       volume = {680},
          eid = {A22},
        pages = {A22},
          doi = {10.1051/0004-6361/202347909},
archivePrefix = {arXiv},
       eprint = {2310.06539},
 primaryClass = {astro-ph.SR},
       adsurl = {https://ui.adsabs.harvard.edu/abs/2023A&A...680A..22M}
}

@ARTICLE{EVO_Pauli2023,
       author = {{Pauli}, D. and {Oskinova}, L.~M. and {Hamann}, W.-R. and {Bowman}, D.~M. and {Todt}, H. and {Shenar}, T. and {Sander}, A.~A.~C. and {Erba}, C. and {G{\'o}mez-Gonz{\'a}lez}, V.~M.~A. and {Kehrig}, C. and {Klencki}, J. and {Kuiper}, R. and {Mehner}, A. and {de Mink}, S.~E. and {Oey}, M.~S. and {Ramachandran}, V. and {Schootemeijer}, A. and {Reyero Serantes}, S. and {Wofford}, A.},
        title = "{Spectroscopic and evolutionary analyses of the binary system AzV 14 outline paths toward the WR stage at low metallicity}",
      journal = {\aap},
         year = 2023,
        month = may,
       volume = {673},
          eid = {A40},
        pages = {A40},
          doi = {10.1051/0004-6361/202345881},
archivePrefix = {arXiv},
       eprint = {2303.03989},
 primaryClass = {astro-ph.SR},
       adsurl = {https://ui.adsabs.harvard.edu/abs/2023A&A...673A..40P}
}

@ARTICLE{SMCMET_Hunter2007,
       author = {{Hunter}, I. and {Dufton}, P.~L. and {Smartt}, S.~J. and {Ryans}, R.~S.~I. and {Evans}, C.~J. and {Lennon}, D.~J. and {Trundle}, C. and {Hubeny}, I. and {Lanz}, T.},
        title = "{The VLT-FLAMES survey of massive stars: surface chemical compositions of B-type stars in the Magellanic Clouds}",
      journal = {\aap},
         year = 2007,
        month = apr,
       volume = {466},
       number = {1},
        pages = {277-300},
          doi = {10.1051/0004-6361:20066148},
archivePrefix = {arXiv},
       eprint = {astro-ph/0609710},
 primaryClass = {astro-ph},
       adsurl = {https://ui.adsabs.harvard.edu/abs/2007A&A...466..277H}
}

@ARTICLE{SMCMET_Korn2000,
       author = {{Korn}, A.~J. and {Becker}, S.~R. and {Gummersbach}, C.~A. and {Wolf}, B.},
        title = "{Chemical abundances from Magellanic cloud B stars}",
      journal = {\aap},
         year = 2000,
        month = jan,
       volume = {353},
        pages = {655-665},
       adsurl = {https://ui.adsabs.harvard.edu/abs/2000A&A...353..655K}
}

@article{WRSURVEY_Hamann2006,
	author = {W. -R. Hamann and G. Graefener and A. Liermann},
	doi = {10.1051/0004-6361:20065052},
	issue = {3},
	journal = {\aap},
	month = {8},
	pages = {1015-1031},
	title = {The Galactic WN stars: Spectral analyses with line-blanketed model atmospheres versus stellar evolution models with and without rotation},
	volume = {457},
	url = {http://arxiv.org/abs/astro-ph/0608078 http://dx.doi.org/10.1051/0004-6361:20065052},
	year = {2006},
}

@article{NGCWR_Armandroff1985A,
	author = {T. E. Armandroff and P. Massey},
	doi = {10.1086/163107},
	issn = {0004-637X},
	journal = {\apj},
	month = {4},
	pages = {685},
	publisher = {American Astronomical Society},
	title = {Wolf-Rayet stars in NGC 6822 and IC 1613},
	volume = {291},
	year = {1985},
}

@article{NGCWR_Bianchi2001,
	author = {Luciana Bianchi and Salvatore Scuderi and Philip Massey and Martino Romaniello},
	doi = {10.1086/319969},
	issn = {00046256},
	issue = {4},
	journal = {\apj},
	month = {4},
	pages = {2020-2031},
	publisher = {American Astronomical Society},
	title = {The Massive Star Content of NGC 6822: Ground-based and[ITAL]HUBBLE SPACE TELESCOPE[/ITAL][ITAL]Hubble Space Telescope[/ITAL] Photometry},
	volume = {121},
	year = {2001},
}

@ARTICLE{TRACKS_pauli2026,
       author = {{Pauli}, D. and {Langer}, N. and {Schootemeijer}, A. and {Marchant}, P. and {Jin}, H. and {Ercolino}, A. and {Picco}, A. and {Willcox}, R. and {Sana}, H.},
        title = "{The drastic impact of Eddington-limit induced mass ejections on massive star populations}",
      journal = {arXiv e-prints},
         year = 2026,
        month = jan,
          eid = {arXiv:2601.08822},
        pages = {arXiv:2601.08822},
          doi = {10.48550/arXiv.2601.08822},
archivePrefix = {arXiv},
       eprint = {2601.08822},
 primaryClass = {astro-ph.SR},
       adsurl = {https://ui.adsabs.harvard.edu/abs/2026arXiv260108822P}
}

@article{WRCLASS_vanderhucht2001,
	author = {Karel A van der Hucht},
	doi = {https://doi.org/10.1016/S1387-6473(00)00112-3},
	issn = {1387-6473},
	issue = {3},
	journal = {New Astronomy Reviews},
	pages = {135-232},
	title = {The VIIth catalogue of galactic Wolf–Rayet stars},
	volume = {45},
	url = {https://www.sciencedirect.com/science/article/pii/S1387647300001123},
	year = {2001},
}

@article{ROT_Shenar2014,
	author = {T. Shenar and W. R. Hamann and H. Todt},
	doi = {10.1051/0004-6361/201322496},
	issn = {0004-6361},
	journal = {\aap},
	month = {2},
	pages = {A118},
	publisher = {EDP Sciences},
	title = {The impact of rotation on the line profiles of Wolf-Rayet stars},
	volume = {562},
	url = {https://www.aanda.org/articles/aa/full_html/2014/02/aa22496-13/aa22496-13.html https://www.aanda.org/articles/aa/abs/2014/02/aa22496-13/aa22496-13.html},
	year = {2014},
}

@ARTICLE{ESOREFLEX_Freduling2013,
	author = {{Freudling}, W. and {Romaniello}, M. and {Bramich}, D.~M. and
	{Ballester}, P. and {Forchi}, V. and {Garc{\'{\i}}a-Dabl{\'o}}, C.~E. and
	{Moehler}, S. and {Neeser}, M.~J.},
	title = "{Automated data reduction workflows for astronomy. The ESO Reflex environment}",
	journal = {\aap},
	archivePrefix = "arXiv",
	eprint = {1311.5411},
	primaryClass = "astro-ph.IM",
	year = 2013,
	month = nov,
	volume = 559,
	eid = {A96},
	pages = {A96},
	doi = {10.1051/0004-6361/201322494},
	adsurl = {http://adsabs.harvard.edu/abs/2013A%26A...559A..96F}
}

@ARTICLE{POWR_Grafner2002,
	author = {{Gr{\"a}fener}, G. and {Koesterke}, L. and {Hamann}, W. -R.},
	title = "{Line-blanketed model atmospheres for WR stars}",
	journal = {\aap},
	year = 2002,
	month = may,
	volume = {387},
	pages = {244-257},
	adsurl = {https://ui.adsabs.harvard.edu/abs/2002A&A...387..244G}
}

@ARTICLE{WRSMC_Shenar2016,
	author = {{Shenar}, T. and {Hainich}, R. and {Todt}, H. and {Sander}, A. and {Hamann}, W. -R. and {Moffat}, A.~F.~J. and {Eldridge}, J.~J. and {Pablo}, H. and {Oskinova}, L.~M. and {Richardson}, N.~D.},
	title = "{Wolf-Rayet stars in the Small Magellanic Cloud. II. Analysis of the binaries}",
	journal = {\aap},
	year = 2016,
	month = jun,
	volume = {591},
	eid = {A22},
	pages = {A22},
	doi = {10.1051/0004-6361/201527916},
	archivePrefix = {arXiv},
	eprint = {1604.01022},
	primaryClass = {astro-ph.SR},
	adsurl = {https://ui.adsabs.harvard.edu/abs/2016A&A...591A..22S}
}

@ARTICLE{WRSMC_Hainich2015,
	author = {{Hainich}, R. and {Pasemann}, D. and {Todt}, H. and {Shenar}, T. and {Sander}, A. and {Hamann}, W. -R.},
	title = "{Wolf-Rayet stars in the Small Magellanic Cloud. I. Analysis of the single WN stars}",
	journal = {\aap},
	year = 2015,
	month = sep,
	volume = {581},
	eid = {A21},
	pages = {A21},
	doi = {10.1051/0004-6361/201526241},
	archivePrefix = {arXiv},
	eprint = {1507.04000},
	primaryClass = {astro-ph.SR},
	adsurl = {https://ui.adsabs.harvard.edu/abs/2015A&A...581A..21H}
}

@misc{Hips2Fits,
	title = {Hips2Fits},
	author = {{Université de Strasbourg/CNRS}},
	year = {2023},
	howpublished={\url{https://alasky.u-strasbg.fr/hips-image-services/hips2fits}},
	note = {Accessed: 2023-03-04}
}

@ARTICLE{Chene2019,
       author = {{Chen{\'e}}, A.-N. and {St-Louis}, N. and {Moffat}, A.~F.~J. and {Schnurr}, O. and {Crowther}, P.~A. and {Wade}, G.~A. and {Richardson}, N.~D. and {Baranec}, C. and {Ziegler}, C.~A. and {Law}, N.~M. and {Riddle}, R. and {Rate}, G.~A. and {Artigau}, {\'E}. and {Alecian}, E. and {BinaMIcS Collaboration}},
        title = "{Investigating the origin of the spectral line profiles of the Hot Wolf-Rayet Star WR 2}",
      journal = {\mnras},
         year = 2019,
        month = apr,
       volume = {484},
       number = {4},
        pages = {5834-5844},
          doi = {10.1093/mnras/stz411},
archivePrefix = {arXiv},
       eprint = {1905.05815},
 primaryClass = {astro-ph.SR},
       adsurl = {https://ui.adsabs.harvard.edu/abs/2019MNRAS.484.5834C}
}

@ARTICLE{Patrick2015,
       author = {{Patrick}, L.~R. and {Evans}, C.~J. and {Davies}, B. and {Kudritzki}, R.-P. and {Gazak}, J.~Z. and {Bergemann}, M. and {Plez}, B. and {Ferguson}, A.~M.~N.},
        title = "{Red Supergiant Stars as Cosmic Abundance Probes: KMOS Observations in NGC 6822}",
      journal = {\apj},
         year = 2015,
        month = apr,
       volume = {803},
       number = {1},
          eid = {14},
        pages = {14},
          doi = {10.1088/0004-637X/803/1/14},
archivePrefix = {arXiv},
       eprint = {1501.07601},
 primaryClass = {astro-ph.SR},
       adsurl = {https://ui.adsabs.harvard.edu/abs/2015ApJ...803...14P}
}

@ARTICLE{NGCWR_Armandroff1991,
       author = {{Armandroff}, Taft E. and {Massey}, Philip},
        title = "{Wolf-Rayet Stars in Local Group Galaxies: Numbers and Spectral Properties}",
      journal = {\aj},
         year = 1991,
        month = sep,
       volume = {102},
        pages = {927},
          doi = {10.1086/115924},
       adsurl = {https://ui.adsabs.harvard.edu/abs/1991AJ....102..927A}
}

@ARTICLE{NGC_Khatamsaz2024,
       author = {{Khatamsaz}, F. and {Abdollahi}, M. and {Abdollahi}, H. and {Javadi}, A. and {van Loon}, J. Th.},
        title = "{Star Formation History of the Local Group Dwarf Irregular Galaxy, NGC 6822}",
      journal = {Communications of the Byurakan Astrophysical Observatory},
         year = 2024,
        month = dec,
       volume = {71},
        pages = {394-397},
          doi = {10.52526/25792776-24.71.2-394},
archivePrefix = {arXiv},
       eprint = {2412.05646},
 primaryClass = {astro-ph.GA},
       adsurl = {https://ui.adsabs.harvard.edu/abs/2024CoBAO..71..394K}
}

@article{WRLMC_hainich2014,
	author = {{Hainich}, R. and {R\"uhling, U.} and {Todt, H.} and {Oskinova, L. M.} and {Liermann, A.} and {Gr\"afener, G.} and {Foellmi, C.} and {Schnurr, O.} and {Hamann, W.-R.}},
	title = {The Wolf-Rayet stars in the Large Magellanic Cloud - A comprehensive analysis of the WN class},
	DOI= "10.1051/0004-6361/201322696",
	url= "https://doi.org/10.1051/0004-6361/201322696",
	journal = {A\&A},
	year = 2014,
	volume = 565,
	pages = "A27",
	month = "",
}

@ARTICLE{lefever2026,
       author = {{Lefever}, R.~R. and {Sander}, A.~A.~C. and {Bernini-Peron}, M. and {Gonz{\'a}lez-Tor{\`a}}, G. and {Hamann}, W.-R. and {Josiek}, J. and {Ramachandran}, V. and {Sch{\"o}sser}, E.~C. and {Todt}, H.},
        title = "{Dynamically consistent analysis of Galactic WN4b stars: Revised parameters and insights on Wolf─Rayet mass-loss treatments}",
      journal = {\aap},
         year = 2026,
        month = feb,
       volume = {707},
          eid = {A8},
        pages = {A8},
          doi = {10.1051/0004-6361/202558211},
archivePrefix = {arXiv},
       eprint = {2601.02498},
 primaryClass = {astro-ph.SR},
       adsurl = {https://ui.adsabs.harvard.edu/abs/2026A&A...707A...8L}
}

\begin{appendix}

\onecolumn
\section{Additional Models} \label{App:models}

\begin{figure*}
	\begin{center}	
		\includegraphics[width=0.85\linewidth]{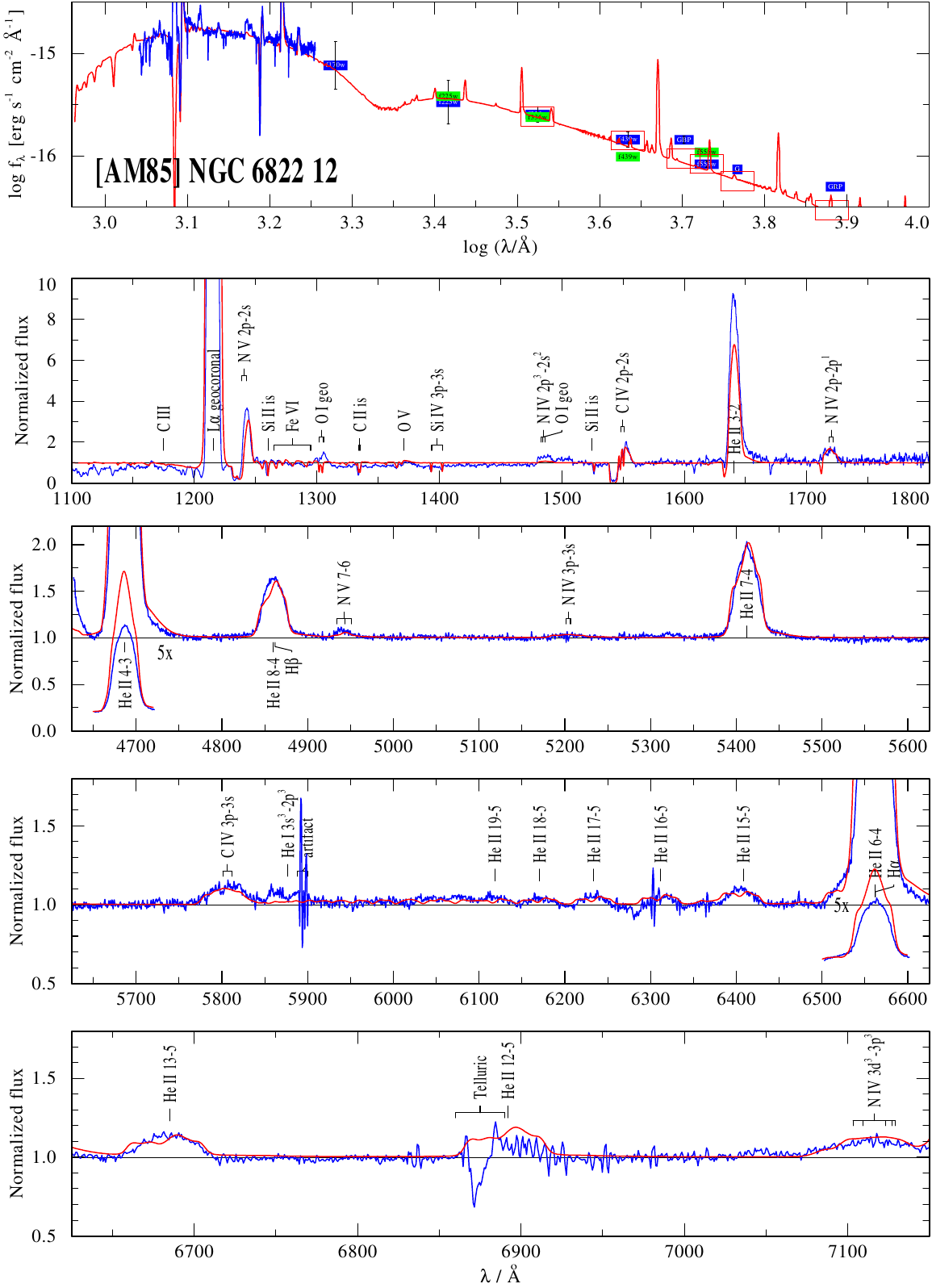}
	\end{center}
    \caption{Comparison between observed and synthetic spectra for the best-fit single star model for star \#12, shown by blue and red lines respectively. In the upper panel, blue boxes indicate photometry extracted from HST images and from Gaia (see Sect.~\ref{Obs_Photometry}). For this star, the additional green boxes indicate photometry by \citet{NGCWR_Bianchi2001}. Red open boxes are centred on synthetic photometry, as calculated from the model flux and the corresponding filter function. The model parameters are given in Table~\ref{tab:Single_parameters}.}
    \label{WR12single_fullmasterplot}
\end{figure*}

\begin{figure*}
	\begin{center}	
		\includegraphics[width=0.9\linewidth]{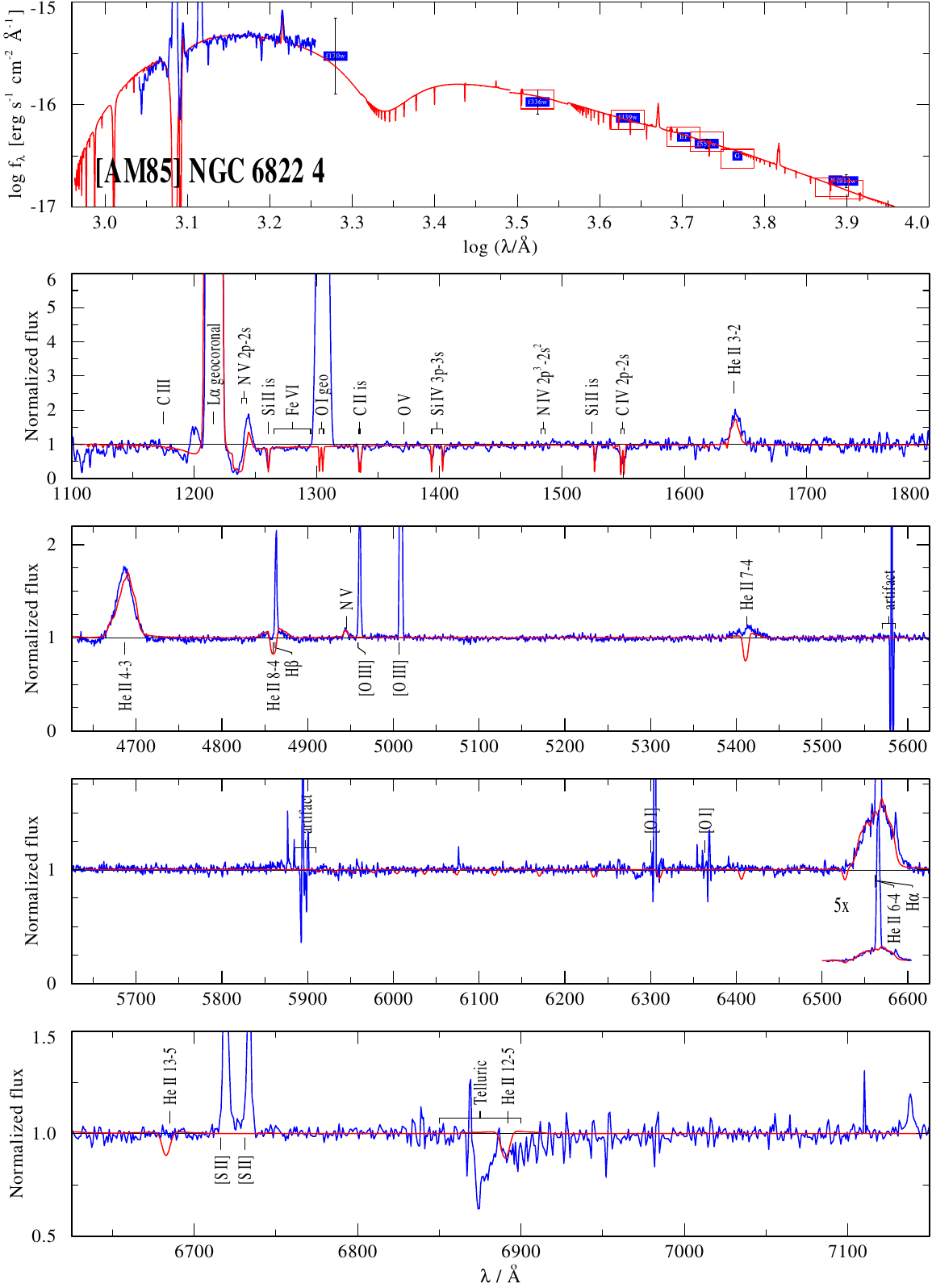}
	\end{center}
    \caption{Same as Fig.\ref{WR12single_fullmasterplot}, but for star \#4. The model parameters are given in Table~\ref{tab:Single_parameters}.}
    \label{WR4single_fullmasterplot}
\end{figure*}

\begin{figure*}
	\begin{center}	
		\includegraphics[width=0.9\linewidth]{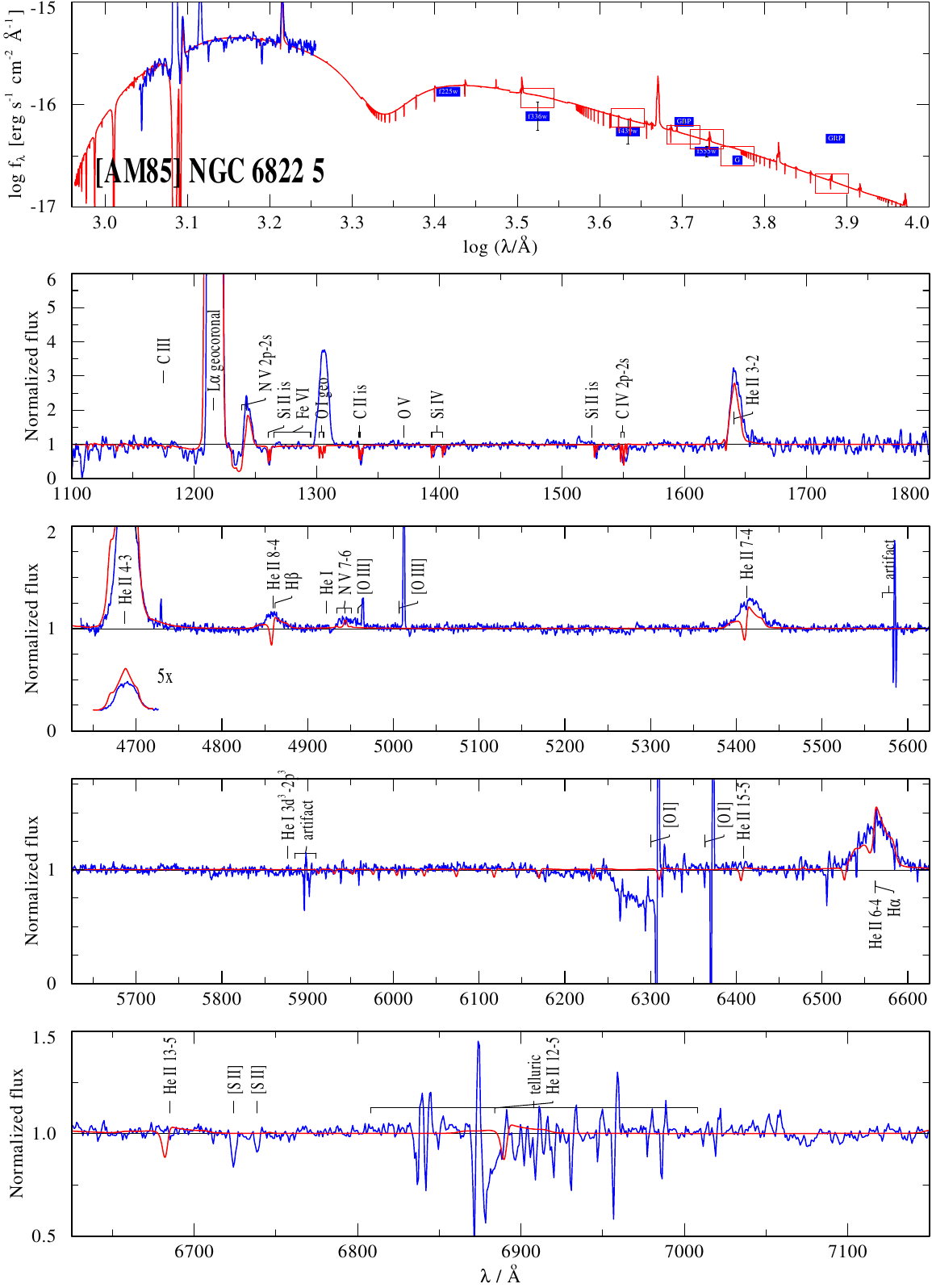}
	\end{center}
    \caption{Same as Fig.~\ref{WR12single_fullmasterplot} but for star \#5. The model parameters are given in Table~\ref{tab:Single_parameters}.}
    \label{WR5single_fullmasterplot}
\end{figure*}

\begin{figure*}
	\begin{center}	
		\includegraphics[width=0.9\linewidth]{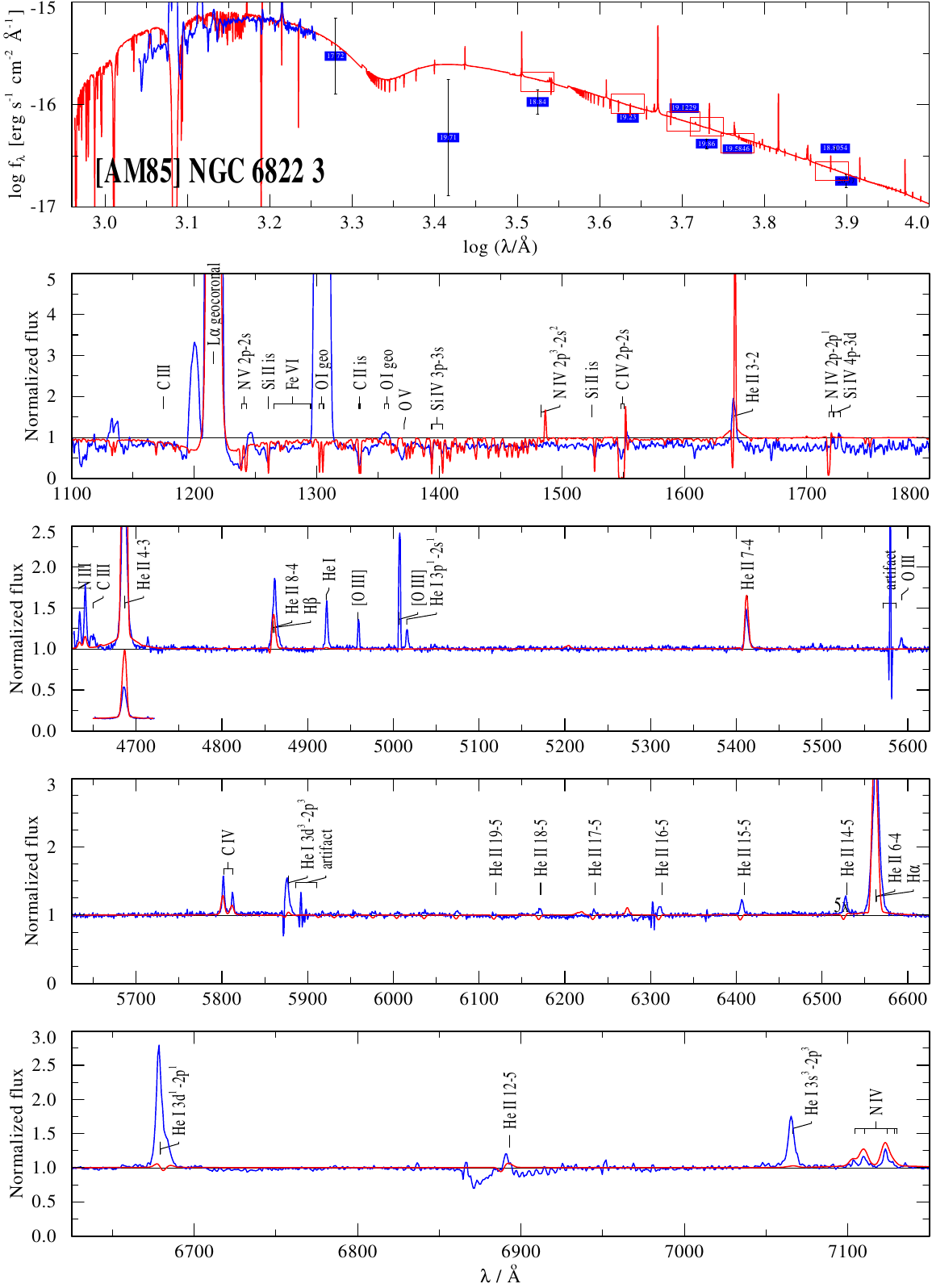}
	\end{center}
    \caption{Same as Fig.~\ref{WR12single_fullmasterplot} but for an example single-star fit for star \#3.}
    \label{WR3single_fullmasterplot}
\end{figure*}

As discussed in Sect.~\ref{Results}, whilst we believe all four stars are more likely binaries, single-star models have also been tested. In this section we present the best-fit single-star models for these stars.

For star \#12, we achieve the best internal consistency between the individual lines with the depth-dependent clumping description from \citet{CLUMPING_Najarro2009}:
\begin{equation}
    f = f_\text{max} + (1 - f_\text{max})\exp\left(\frac{\varv(r)}{\varv_{\text{cl}, 1}}\right) + (f_\infty - f_\text{max})\exp\left(-\frac{\varv_\infty - \varv(r)}{\varv_{\text{cl}, 2}}\right)
\end{equation}
Values of $D_\infty = f_\infty^{-1} = 4$, $\varv_{\text{cl}, 1} = 1$, and $\varv_{\text{cl}, 2} = 150$ are found to be suitable. The maximum value of the clumping factor is set to $D_\text{max} = f_\text{max}^{-1}= 10$, in order to reproduce the shape of the electron scattering wings. 

\begin{table*}[h!]\centering
	\caption{Summary of model parameters of the best-fit single-star models for stars \#4, \#5, and \#12.}
		\begin{tabular}{cccc}
			\hline\hline
			Properties/Stars  & Star \#4 & Star \#5 & Star \#12\\
			\hline
			
			$E(B-V)$ [mag]	 & 0.38$\pm0.005$ &  0.39$\pm0.005$ & 0.29$\pm0.005$ \\
			
			$T_\ast$ [kK] & 84$\pm10$ & 90$\pm10$ & 85$\pm10$\\

            $T_{2/3}$ [kK] & 82 & 78 & 84 \\

            $\log(L)$ [L$_\odot$] & 6.00$\pm$0.2 & 6.1$\pm$0.2 & 6.15$\pm$0.2 \\
			
			$\log{R_\text{t}}~[R_\odot]$  &1.20$\pm$0.05 & 1.31$\pm$0.05 & 1.05$\pm$0.05 \\
			
			$\log(\dot{M})$  [M$_\odot$yr$^{-1}$] & -5.45$\pm$0.2& -5.1$\pm$0.2 & -4.95$\pm$0.2 \\

            $\log(\dot{M_t})$ & -5.353 & -4.603 & -4.976 \\

            $\varv_{\text{rad},\ast}$ [km s$^{-1}$] & -100$\pm$20 & -350$\pm$50 & -50$\pm$50 \\

            $\varv_{\infty}$ [km s$^{-1}$] & 1600$\pm$100 & 1800$\pm$100 & 1600$\pm$100 \\
            
			$M_V$ [mag]  & -4.73$\pm0.3$ & -5.21$\pm0.3$ & -5.32$\pm0.3$ \\

			$X_\text{H}$ & 0.4$\pm0.05$ & 0$^{+0.05}$ & 0.05$\pm0.02$ \\

            $X_\text{Fe}/10^{-4}$ & 3 & 3 & 1$\pm0.5$ \\
            
            $X_\text{C}/10^{-5}$ & 2.5$\pm2$ & 1.5$\pm1$ & 2.5$\pm1$ \\
            
            $X_\text{N}/10^{-3}$ & 1.5$\pm0.5$ & 1.5$\pm0.2$ & 1.5$\pm0.2$ \\

            $X_\text{O}/10^{-5}$ & 2.5 & 2.5 & 2.5 \\

			$D_\infty$ & 4 & 4 & 10$\rightarrow$4 \\
            
			$R$ [$R_\odot$] & 4.7$^{+1.8}_{-1.4}$ & 4.6$^{+1.7}_{-1.3}$ & 5.5$^{+2.1}_{-1.6}$\\

            $M$ [$M_\odot$] & 31.1$^{+18.2}_{-11.5}$ & 37.1$^{+21.7}_{-13.7}$ & 40.1$^{+23.5}_{-14.5}$ \\

            $\log{Q_\text{H}}$~[s\textsuperscript{-1}] & 49.91 & 49.06 & 50.02 \\
            
            $\log{Q_\text{He\,\scshape I}}$~[s\textsuperscript{-1}] & 49.70 & 48.83 & 49.77 \\
            
            $\log{Q_\text{He\,\scshape II}}$~[s\textsuperscript{-1}]  & 47.77 & 39.13 & ...$^c$\\

			\hline
            
		\end{tabular}		
    \tablefoot{$^\text{c}$ Estimated using the mass-luminosity relations from \citet{MASSLUM_Grafener2011}. $^\text{c}$ optically thick at outer boundary for He\,{\scshape ii} ionizing photons.}
	\label{tab:Single_parameters}
\end{table*}

\begin{table}[H]
\begin{subtable}[t]{0.4\textwidth}\centering
	\caption{Summary of parameters for the closest single-star track models to stars \#4, \#5, and \#12 from \citet{TRACKS_pauli2026}.}
		\begin{tabular}{cccc}
			\hline\hline
			Properties/Stars & Star \#4 & Star \#5 & Star \#12 \\
			\hline
            Initial Mass ($M_i$) & 50.2 & 112 & 112 \\
            
			Age (Myr) & 4.5 & 3.0 & 3.0 \\
			
			Mass ($M_\odot$) & 29.1 & 37.5 & 37.5 \\
			
			$T$ [~kK] & 72.2 & 93.3 & 85.5 \\

            $\log(L)$ [$L_\odot$] & 6.98 & 6.08 & 6.07 \\
			
			$\log(\dot{M})$ [$M_\odot$yr$^{-1}$] & -5.18 & -4.23 & -4.24 \\

            $X_\text{H}$ & 0.27 & 0.01 & 0.01\\
            
            $X_\text{C}/10^{-5}$ & 2.1 & 3.7 & 3.7 \\
            
            $X_\text{N}/10^{-3}$ & 1.6 & 1.5 & 1.5 \\
            
            $X_\text{O}/10^{-5}$ & 1.8 & 1.2 & 1.2 \\

			\hline
            
		\end{tabular}
	\label{tab:Model_parameters_app}
\end{subtable}\hspace{5mm}
\begin{subtable}[t]{0.5\textwidth}\centering
	\caption{Summary of parameters for the closest BPASS binary track models to stars \#4, \#5, and \#12.}
		\begin{tabular}{cccc}
			\hline\hline
			Properties/Stars & Star \#4 & Star \#5 & Star \#12 \\
			\hline
            Initial Mass ($M_{i, 1}$) & 40 & 25 & 21 \\

            Initial Mass Ratio $q_i$ & 0.5 & 0.5 & 0.9 \\

            Initial Orbital Period $\log P_i$ [days] & 1.0 & 0.4 & 0.8 \\
            
			Age (Myr) & 5.37 & 8.35 & 10.23 \\

			Mass ($M_\odot$) & 16.8 & 8.5 & 8.1 \\

            Orbital Period $\log P$ [days] & 0.60 & 0.46 & 0.44 \\
            
			$T$ [~kK] & 84 & 122 & 86 \\

            $\log(L)$ [$L_\odot$] & 5.65 & 5.19 & 5.84 \\

            $X_\text{H}$ & 0.25 & 0.03 & 0.1 \\
            
            $X_\text{C}/10^{-5}$ & 1.3 & 1.5 & 1.7 \\
            
            $X_\text{N}/10^{-3}$ & 1.3 & 1.3 & 1.3 \\
            
            $X_\text{O}/10^{-5}$ & 5.3 & 2.9 & 2.3 \\

			$T_2$ [~kK] & 39 & 36 & 33 \\

            $\log(L_2)$ [$L_\odot$] & 4.8 & 4.2 & 4.9 \\

			\hline
            
		\end{tabular}
	\label{tab:Model_parameters}
\end{subtable}
\end{table}

\begin{figure*}
	\begin{center}	
		\includegraphics[width=0.9\linewidth]{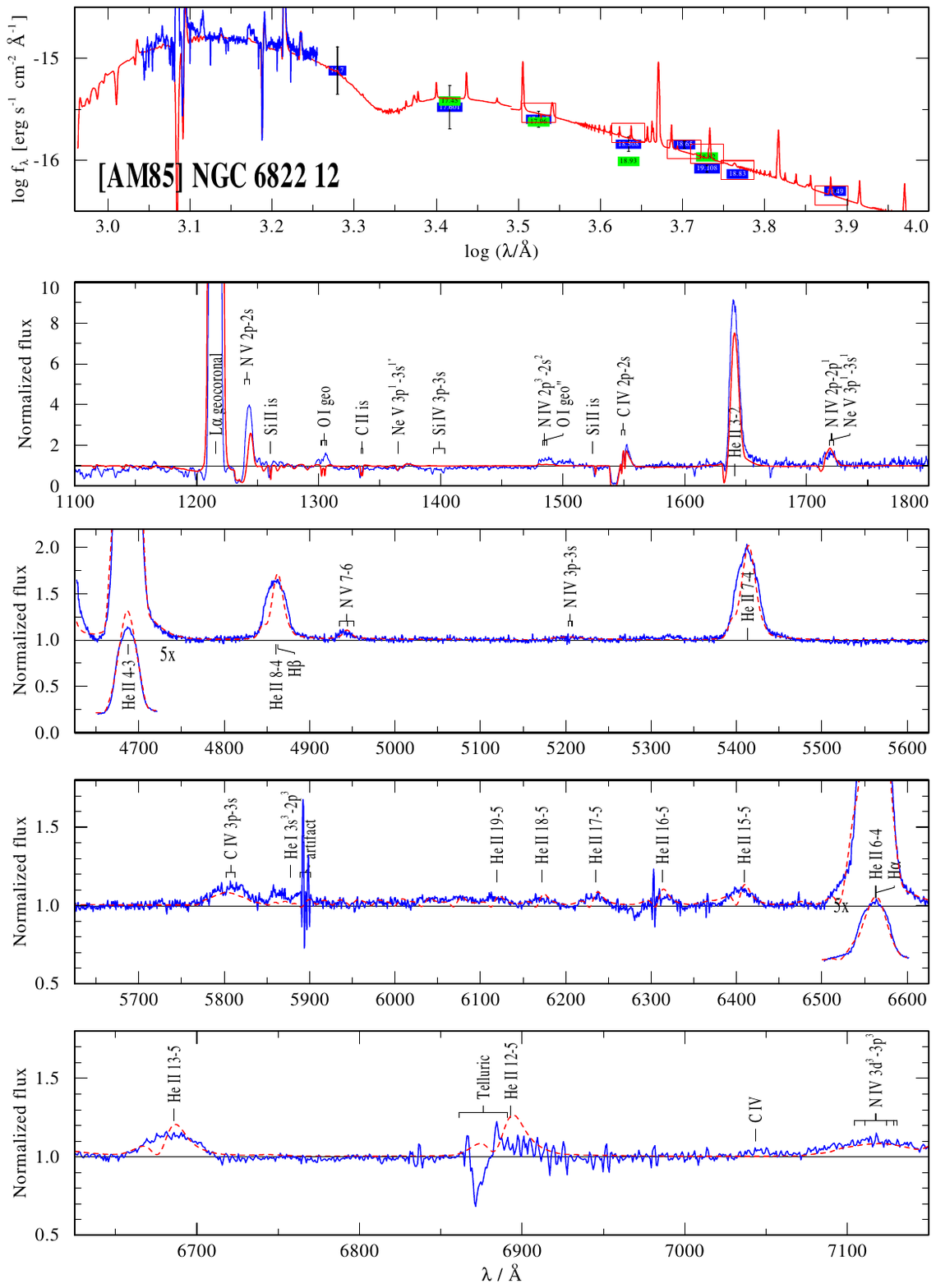}
	\end{center}
    \caption{Same as Fig.\ref{WR12single_fullmasterplot}, but using a standard beta-law velocity field with $\beta=1$.}
    \label{WR12_NOVELOfullmasterplot}
\end{figure*}

\twocolumn
\begin{figure}[h!]
	\begin{center}	
		\includegraphics[width=0.9\linewidth]{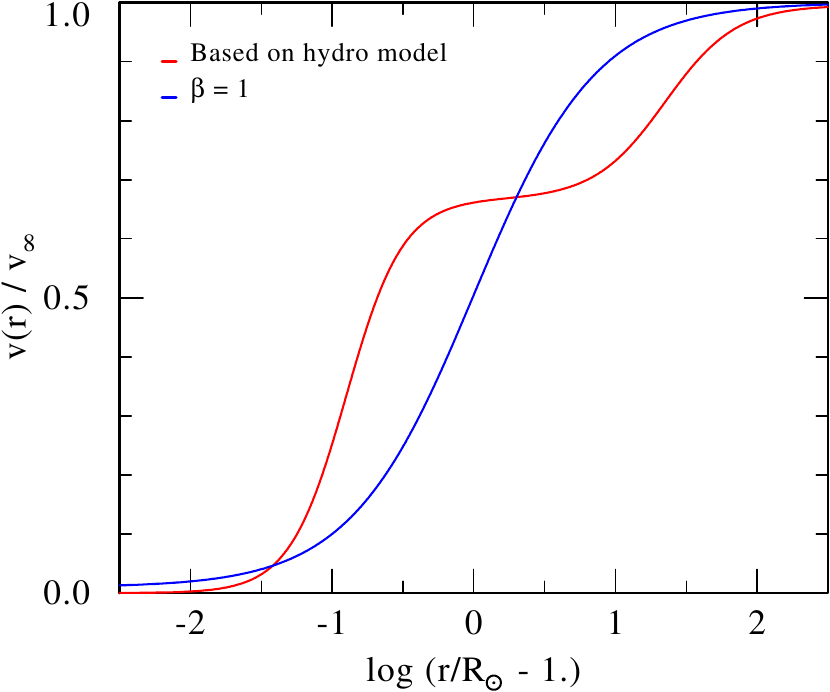}
	\end{center}
    \caption{The velocity field from hydrodynamically consistent models, used for single-star models for stars \#4, \#5, and \#12 (shown in Figures~\ref{WR5_fullmasterplot}, \ref{WR12_fullmasterplot}, and \ref{WR4_fullmasterplot}).}
    \label{Velocity field}
\end{figure}

\section{Evolutionary Tracks for Single-star Models} \label{app:evo}

The observed positions in the HR diagram of the single-star fits for stars \#4, \#5, and \#12 are compared with new evolutionary tracks for single stars at SMC metallicity from \citet{TRACKS_pauli2026}.

\begin{figure}[h]
	\begin{center}	
		\includegraphics[width=\linewidth]{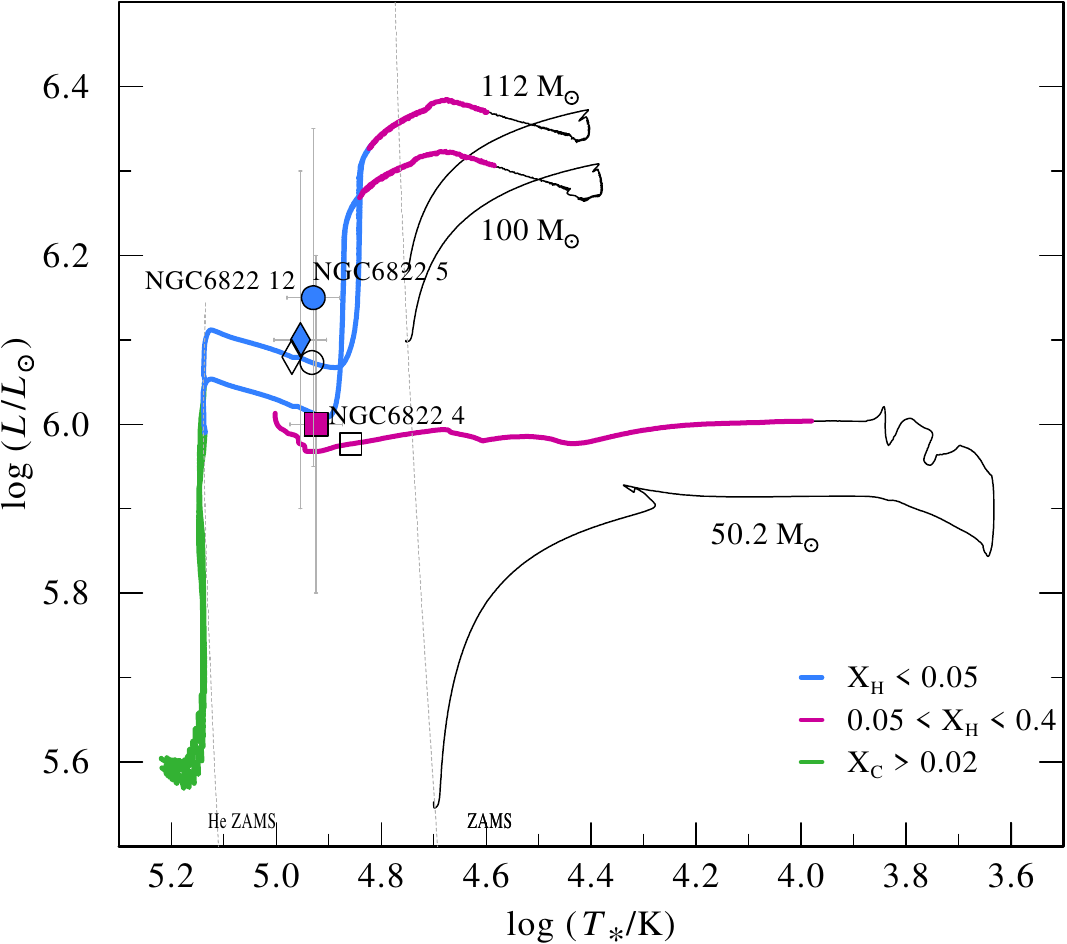}
	\end{center}
    \caption{The positions of stars \#4, \#5, and \#12 in the HR diagram. Note that we identify the stellar temperature $T_\ast$ (see Sect.~\ref{Stellar Atmosphere Modelling}) with the effective temperature as used in the evolutionary models. Stellar evolutionary tracks from \citet{TRACKS_pauli2026}, with an SMC-like composition, are shown. The selected tracks have initial masses of 112~$M_\odot$ and 50.2~$M_\odot$, and are shown as black and coloured lines. The colour of the lines and markers indicate the surface hydrogen abundance according to the legend. Symbols are matched for the star and its corresponding model, with the hollow symbols indicating the closest model on the tracks (see text)}
    \label{hrd+tracks_app}
\end{figure}

For each star in our sample, we look for a model with age $t$ on a track defined by its initial mass $M_i$ which reproduces the values of $T_\ast$, $L$, and $X_H$ as well as possible. This is done with a $\chi^2$ fitting procedure, using the values of $\log T_\ast$, $\log L$, and $X_\text{H}$ for each star. The tracks are sufficiently dense in time such that no interpolation between ages is necessary. We do not interpolate between the different mass-defined tracks.

The best-fit tracks are shown in an HR diagram in Fig.~\ref{hrd+tracks_app}, with stellar parameters listed in Table~\ref{tab:Model_parameters_app}. Stars \#5 and \#12 are best reproduced by the same evolutionary track with an initial mass of $M_i = 112~$M$_\odot$, whilst star \#4 is well matched by the track with $M_i = 50.2$~M$_\odot$. The best-fit track models are within 2$\sigma$ of most stellar parameters. Our analysis clearly illustrates all three stars are consistent with single-star evolution. Based on the best-matching tracks, we predict the single-star solutions for \#5 and \#12 would likely finally collapse into fairly heavy massive black holes. With such high masses, these stars would be expected to have formed in massive clusters.

The observed mass-loss rates (Table~\ref{tab:Single_parameters}) are noticeably lower than those given by the mass loss prescription adopted for the tracks (Table~\ref{tab:Model_parameters_app}). For star \#4, this likely results in the difference between predicted and observed hydrogen abundances as the outer envelope is more efficiently stripped. 

\begin{table*}[h!]\centering

\section{Tables of Observations} \label{app:obs}

    \caption{Log of optical observations, made with FORS~2 at the VLT.}
    	\begin{tabular}{ccccccc}
    		\hline\hline
    		Object & Date (YYYY-MM-DD) & Grism & $\lambda$ (\AA) & Start Time (UT)\rule[0mm]{0mm}{4mm}\\
            \hline
            {[AM85]} NGC~6822 3 & 2022-05-31 & GRIS\_1200R$+$93 & 5750 - 7310 & 07:35:20\\
            & 2022-05-31 & GRIS\_1200R$+$93 & 5750 - 7310 & 06:38:52\\
            & 2022-07-02 &GRIS\_1400V$+$18 & 4560 - 5860 & 03:13:58\\
            & 2022-07-01 &GRIS\_1400V$+$18 & 4560 - 5860 & 04:28:24\\
    		\hline
            {[AM85]} NGC~6822 4 & 2022-04-28 & GRIS\_1200R$+$93 & 5750 - 7310 & 08:20:19\\
            & 2022-05-26 & GRIS\_1200R$+$93 & 5750 - 7310 & 08:17:55\\
            & 2022-05-28 &GRIS\_1400V$+$18 & 4560 - 5860 & 06:30:20\\
            & 2022-05-28 &GRIS\_1400V$+$18 & 4560 - 5860 & 05:29:42\\
            \hline
            {[AM85]} NGC~6822 5 & 2022-08-22 & GRIS\_1200R$+$93 & 5750 - 7310 & 04:30:44\\
            & 2022-08-24 & GRIS\_1200R$+$93 & 5750 - 7310 & 03:18:27\\
            & 2022-08-24 &GRIS\_1400V$+$18 & 4560 - 5860 & 00:57:06\\
            & 2022-08-24 &GRIS\_1400V$+$18 & 4560 - 5860 & 01:52:38\\
            \hline
            {[AM85]} NGC~6822 12 & 2022-04-28 &GRIS\_1400V$+$18 & 4560 - 5860 & 07:24:44\\
            & 2022-04-28 &GRIS\_1400V$+$18 & 4560 - 5860 & 05:59:36\\
            & 2022-05-01 & GRIS\_1200R$+$93 & 5750 - 7310 & 07:48:11\\
            & 2022-05-04 & GRIS\_1200R$+$93 & 5750 - 7310 & 07:33:58\\
            \hline
            
    	\end{tabular}
    \label{Obs_optical}
\end{table*}

\begin{figure}
	\begin{center}
		\includegraphics[width=\linewidth]{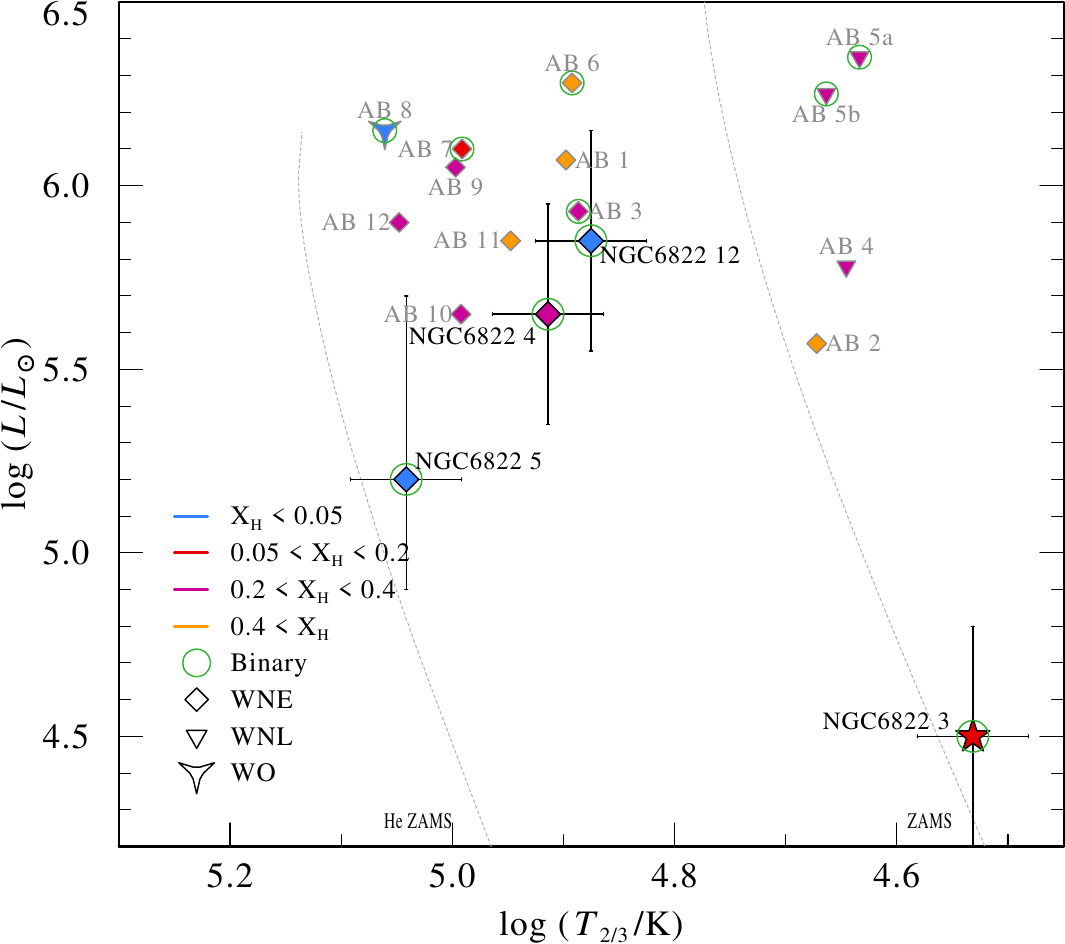}
	\end{center}
    \caption{Same as Fig.\ref{hrd+smc} but with the x-axis showing $T_{2/3}$.}
    \label{hrd+smc_t23}
\end{figure}

\begin{table*}\centering
    \caption{Log of the HST observations analysed in this paper.}
    	\begin{tabular}{cccccccccc}
    		\hline\hline
    		Object & OBS-ID & Date (YYYY-MM-DD) & Start Time (UT) & Exposure (s)\rule[0mm]{0mm}{4mm}\\
    		\hline
            {[AM85]} NGC~6822 3 & lfgt01frq & 2025-03-31 & 18:05:02 & 2762\\
            & & 2025-03-31 & 19:34:23 & 3032\\
            & & 2025-03-31 & 21:08:49 & 2998\\
            & & 2025-03-31 & 22:43:16 & 3150\\
            \hline
            {[AM85]} NGC~6822 4 & lfgt02wgd & 2025-05-23 & 15:28.36 & 2792\\
            & & 2025-05-23 & 16:56:33 & 3098\\
            & & 2025-05-23 & 18:30:53 & 3098\\
            & & 2025-05-23 & 20:05:13 & 3244\\
            \hline
            {[AM85]} NGC~6822 5 & lfgt03k6q & 2025-06-25 & 09:35:51 & 2792\\
            & & 2025-06-25 & 11:05:04 & 3098\\
            & & 2025-06-25 & 12:39:23 & 3098\\
            & & 2025-06-25 & 14:13:43 & 3244\\
            \hline
            {[AM85]} NGC~6822 12 & lfgt04okq & 2025-06-26 & 12:19:49 & 2792\\
    		& & 2025-06-26 & 13:48:32 & 3098\\
            & & 2025-06-26 & 15:22:51 & 3098\\
            & & 2025-06-26 & 16:57:10 & 3244\\
            \hline
            
    	\end{tabular}
    \label{tab:Obs_UV}
\end{table*}

\end{appendix}
\end{document}